\documentclass[12pt,preprint]{aastex}
\usepackage{amsmath,bm}
\usepackage{natbib}
\begin{document}
\slugcomment{Accepted to ApJ 2019-12-30}

\title{X-Ray Ionization of Planet-Opened Gaps in Protostellar Disks}

\author{S.~Y.\ Kim\altaffilmark{1,2,3},
  N.~J.\ Turner\altaffilmark{4}}

\altaffiltext{1}{Division of Physics, Mathematics \& Astronomy,
  California Institute of Technology, Pasadena, California 91125, USA}
\altaffiltext{2}{Department of Astronomy, Ohio State University, 4055
  McPherson Laboratory, 140 West 18th Avenue, Columbus, Ohio 43210,
  USA}
\altaffiltext{3}{Department of Physics, University of Surrey,
  Guildford, Surrey GU2 7XH, United Kingdom}
\altaffiltext{4}{Jet Propulsion Laboratory, California Institute of
  Technology, Pasadena, California 91109, USA;
  neal.turner@jpl.nasa.gov}

\begin{abstract}
  Young planets with masses approaching Jupiter's have tides strong
  enough to clear gaps around their orbits in the protostellar disk.
  Gas flow through the gaps regulates the planets' further growth and
  governs the disks' evolution.  Magnetic forces may drive that flow
  if the gas is sufficiently ionized to couple to the fields.  We
  compute the ionizing effects of the X-rays from the central young
  star, using Monte Carlo radiative transfer calculations to find the
  spectrum of Compton-scattered photons reaching the planet's
  vicinity.  The scattered X-rays ionize the gas at rates similar to
  or greater than the interstellar cosmic ray rate near planets the
  mass of Saturn and of Jupiter, located at 5~au and at 10~au, in
  disks with the interstellar mass fraction of sub-micron dust and
  with the dust depleted a factor 100.  Solving a gas-grain
  recombination reaction network yields charged particle populations
  whose ability to carry currents is sufficient to partly couple the
  magnetic fields to the gas around the planet.  Most cases can
  undergo Hall shear instability, and some can launch
  magnetocentrifugal winds.  However the material on the planet's
  orbit has diffusivities so large in all the cases we examine, that
  magneto-rotational turbulence is prevented and the non-ideal terms
  govern the magnetic field's evolution.  Thus the flow of gas in the
  gaps opened by the young giant planets depends crucially on the
  finite conductivity.
\end{abstract}

\keywords{protoplanetary disks --- X-rays --- radiative transfer ---
  magnetic fields}

\section{INTRODUCTION\label{sec:introduction}}

A protoplanet that grows past the mass of Saturn has gravity strong
enough for its tides to clear an annular gap of low surface density
around its orbit in the protostellar disk, as reviewed by
\citet{2014prpl.conf..667B}.  The planet can continue to grow only if
material from the gap walls is able to reach its Hill sphere.
Furthermore, the planet's subsequent orbital evolution is governed by
the distribution of gas across the gap, especially the amount of
material located near orbital resonances.  The orbit's evolution also
depends on the rate at which disk material crosses the gap.  In this
contribution we investigate whether magnetic forces can act on the
gap.  We explore whether the gas in the gap is ionized well enough to
couple to magnetic fields, so that the fields can displace the
material near the planet.

Young stars emit ionizing X-rays with temperatures of thousands of
electron volts \citep{1999ARA&A..37..363F}, able to penetrate the
circumstellar gas and dust to columns of order 10~g~cm$^{-2}$
\citep{1997ApJ...480..344G, 2013MNRAS.436.3446E}.  By comparison, the
minimum-mass solar nebula has a surface density at 5~au of about
$\Sigma=150$~g~cm$^{-2}$ \citep{1977Ap&SS..51..153W,
  1981PThPS..70...35H}.  Few stellar X-ray photons thus reach the
midplane before planets have grown.  However once a gap opens in the
disk, the X-ray flux at the midplane can increase.  The gas making up
the gap's inner rim will forward-scatter some of the photons arriving
from the star, deflecting a fraction down to the planet's vicinity.
The gap's outer rim, exposed directly to the optical light from the
central star, heats and expands vertically
\citep{2012ApJ...748...92T}, intercepting extra X-ray photons, some of
which will be scattered backward and down into the gap.  A planet
opening a gap thus receives more X-rays than a non-gap-opening planet
at the same location.

Here we examine whether a gap can increase the X-ray flux enough to
contribute significantly to the ionization of gas near the planet.  We
compute the X-ray intensity in the planet's vicinity using a Monte
Carlo radiative transfer approach, estimate the resulting ionization
state by integrating a simple ionization-recombination reaction
network to equilibrium, and compute the plasma's magnetic diffusivity.
We compare against the threshold diffusivities required for the
operation of three mechanisms proposed to drive the accretion flow:
magneto-rotational turbulence, Hall-shear instability, and
magneto-centrifugal winds.  We extend work by
\cite{2015MNRAS.451.1104K} in carrying out X-ray transfer calculations
rather than using results from a gapless disk, while making relatively
simple assumptions about the gap structure and the strength and
geometry of the magnetic fields.

The paper is laid out as follows.  The model star and disk are
described in \S\ref{sec:disk}, the X-ray transfer methods in
\S\ref{sec:transfer}.  The ionization-recombination chemistry and how
we translate the charged species' abundances into diffusivities and
magnetic stresses are set out in \S\ref{sec:chemistry}.  The resulting
X-ray spectra, ionization levels, and magnetic coupling are shown in
\S\ref{sec:results}.  Discussion and conclusions follow in
\S\ref{sec:conclusions}.


\section{STAR AND DISK\label{sec:disk}}

\begin{figure}[tb!]
\centering
\includegraphics[scale=0.75]{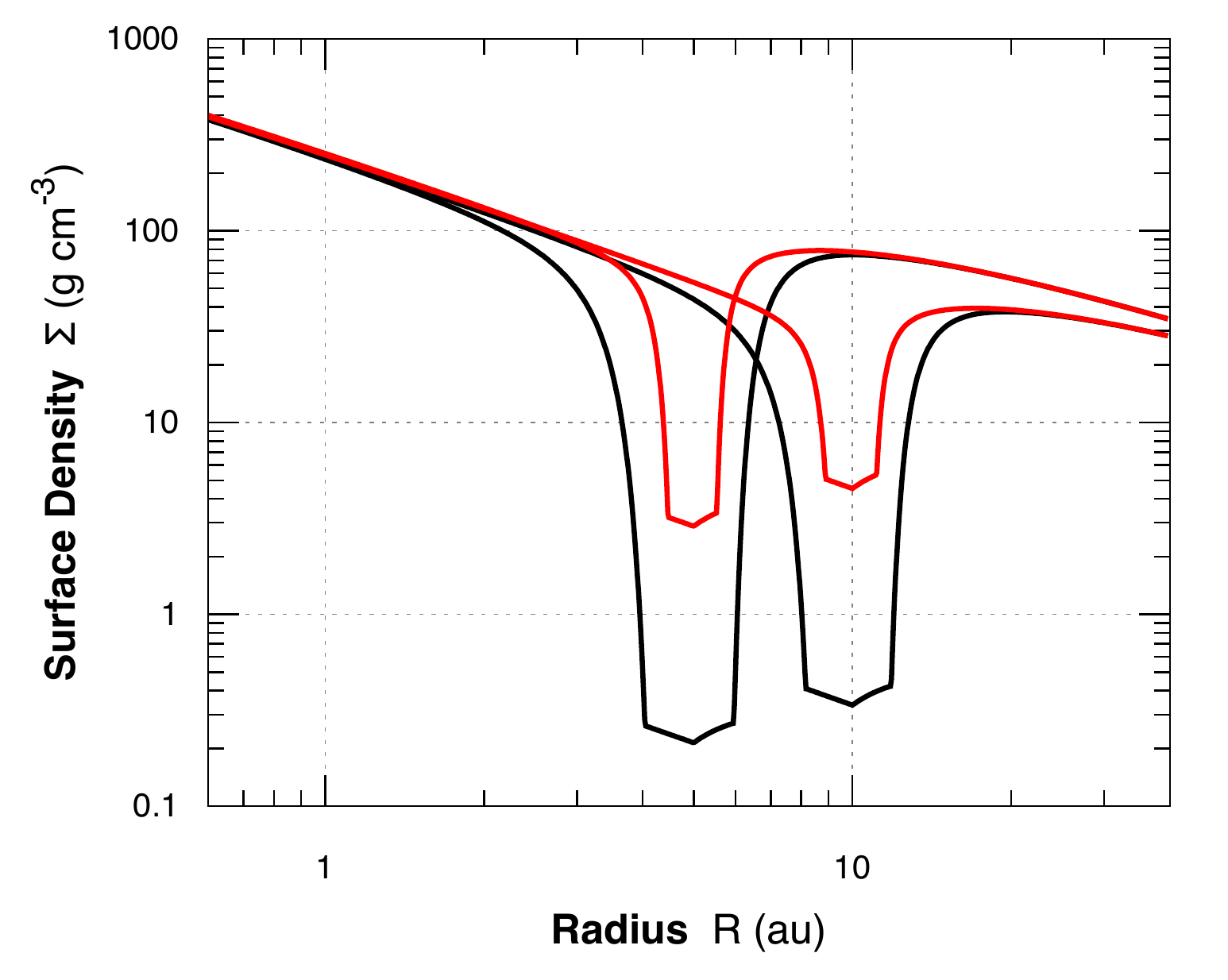}
\caption{Surface density profiles in the four model protostellar
  disks, each with a planet the mass of Saturn (red) or Jupiter
  (black) embedded at 5 or 10~au.}
\label{fig:gap_sd}
\end{figure}

Each model consists of a star, disk, and planet.  The young star is of
solar mass and twice solar radius, and emits a blackbody spectrum with
effective temperature 4500~K.  The resulting luminosity of
$L_*=1.5$~times solar is in the range indicated by stellar evolution
modeling at ages 1--2~Myr \citep{1994ApJS...90..467D,
  2000A&A...358..593S}.  The star is surrounded by an axisymmetric
disk laid down on a radiative transfer grid logarithmically-spaced in
radius with 120~cells from $10^{-1.4}$ to $10^{1.6}$~au (approximately
0.04 to 40~au).  The vertical structure is resolved by dividing each
annulus into 60~cells spaced uniformly from the midplane up to
6~initial pressure scale heights.

We adopt disk surface density profiles from a one-dimensional analytic
model constructed by \cite{2006ApJ...641..526L}, in which the planet's
gap-opening tidal torques balance the disk's gap-closing viscous
stresses \citep{1986ApJ...309..846L}.  The planet follows a fixed
circular orbit.  The accretion stresses within the disk are modeled
using the Shakura-Sunyaev viscous prescription with $\alpha=0.005$,
and the flow is steady-state.

Before any planet is added, the disk's surface density varies
inversely with radius, and is 280~g~cm$^{-2}$ at 1~au.  The planet
partly dams the disk's inflow and accretes most of the material
reaching its orbit.  The ratio of the planet's accretion rate to the
flow rate at the same place in the planet-free disk is $E=6$.  This is
lower than the $E=8$ fiducial case of \cite{2006ApJ...641..526L}
because our model disks have smaller aspect ratios $H/r = 0.034$ to
$0.040$ near the planet.  Other parameters take the fiducial values
from \citet{2006ApJ...641..526L}.

The analytic model breaks down close to the planet where horseshoe
orbits are important, leading to underestimated densities.  In the gap
we therefore set densities using the fit to hydrodynamical results by
\citet[their eq.~12]{2014ApJ...782...88F}.  The surface densities in
our gaps thus depend on three quantities: $\alpha$, the planet-to-star
mass ratio, and the disk aspect ratio.  The gap depths we use are
similar to those obtained including the planet's orbital migration
\citep{2015A&A...574A..52D} and in three-dimensional calculations
\citep{2016ApJ...832..105F}.  However, we note that our disk aspect
ratio of 0.034 at 5~au lies outside the range~0.04 to~0.1 covered by
these three hydrodynamical works.

We place the planet at either $r_p=5$ or 10~au, approximating the
orbits of Jupiter and Saturn, respectively.  To test how magnetic
forces can be expected to vary as the planet grows, at each location
we consider bodies of both Saturn and Jupiter masses, using
planet-to-star mass ratios $q=3\times 10^{-4}$ and $10^{-3}$,
respectively.  For the initial midplane temperature profile we adopt
$T(r) = 124 (L_*/L_\odot)^{1/4} (r/\text{au})^{-1/2}$~K, based on
similar radiative transfer calculations without a gap
\citep{2012ApJ...748...92T}.  This is cooler than the minimum-mass
solar nebula because it includes the effects of the disk's large
optical depth.  The resulting surface density profiles are shown in
figure~\ref{fig:gap_sd}.  The gap in all cases has a high enough
surface density, thanks to the floor from \cite{2014ApJ...782...88F},
that the radial flow speed $v_r={\dot M}/(2\pi r\Sigma)$ needed to
supply the accretion rate is subsonic.

The mass of disk material on the grid ranges from 0.017~$M_\odot$ with
a Jupiter at 10~au, to 0.027~$M_\odot$ with a Saturn at 5~au.  These
are comparable to or a little greater than the minimum-mass solar
nebula.  We verified that the model disks are not susceptible to
fragmentation under their own self-gravity, having the Toomre $Q$
parameter $c_s\Omega/(\pi G\Sigma)$ greater than unity everywhere.
Here $c_s$ is the sound speed, $\Omega$ the orbital frequency, $G$ the
gravitational constant, and $\Sigma$ the surface mass density.  The
mass accretion rate in all models is within 1\% of $3.3\times
10^{-8}$~M$_\odot$~yr$^{-1}$ outside the planet's orbit, and $5\times
10^{-9}$~M$_\odot$~yr$^{-1}$ inside.

The gap modifies the disk's temperature profile from our initial
guess.  Evacuating the gap allows the visible light from the central
star to directly strike the top of the gap's outer wall.  The
starlight heats the wall, increasing its internal gas pressure, so in
hydrostatic equilibrium the wall becomes taller and intercepts yet
more starlight.  The tall wall is likely to intercept more of the
stellar X-rays too.  We therefore include these effects, using an
iterative procedure similar to \citet{2012ApJ...748...92T}.  Given a
density distribution, we obtain new temperatures under radiative
balance with the optical starlight using Monte Carlo transfer with the
\citet{2001ApJ...554..615B} relaxation method.  We then displace gas
up or down to restore vertical hydrostatic balance, holding fixed the
variation of temperature with column.  The new density distribution
serves as the input for the next iteration.  We quit after five
iterations when the structure no longer changes significantly.

The disk's opacity to the optical starlight and reprocessed infrared
radiation comes from dust grains.  We take dust opacities from
\cite{1993A&A...279..577P}, where the particle size distribution
\begin{equation}
n(a) \propto a^{-p}, \qquad a_{min} < a < a_{max},
\end{equation}
with $p = 3.5$.  At temperatures below 125~K, the grains are composed
of a silicate core and an icy mantle whose radius is 14\% of the
core's.  The mantle is polluted with tiny amorphous carbon grains
($a_{min} = 0.007 \mu$m, $a_{max} = 0.03 \mu$m).  Above 125~K, the icy
mantle sublimates, baring a silicate grain ($a_{min} = 0.04 \mu$m,
$a_{max} = 1 \mu$m) and releasing the amorphous carbon grains; the
silicates sublimate at 1500~K, the carbon grains at 2000~K.  The
grains are well-mixed in the gas, and we assume the scattering that
makes up part of the starlight opacity is isotropic.  To model small
grains' incorporation into larger bodies, which is likely to have
occurred by the time planets grow to Saturn or Jupiter mass, we also
compute models with the dust opacities multiplied by a factor
$\epsilon=10^{-2}$.
Considering two values each for the planet
location, the planet mass, and the dust abundance, we compute the
eight model disks listed in table~\ref{tab:models}.  Each model is
given a name whose first digits are the planet location in au,
followed by a letter indicating the planet has Saturn (S) or Jupiter
(J) mass, and a final digit that is the logarithm of the factor by
which the dust abundance is reduced.

\begin{table}[tb!]
\centering
\caption{X-Ray Ionization Rates at the Planet in the Eight Models\label{tab:models}}
\begin{tabular}{rrr ll ll} \hline
Model& Radius   & Mass      & Dust                 & Column         & Ionization Rate          &  Ratio to \\
     & $r_p$, au& Ratio $q$ & Depletion $\epsilon$ & g$\,$cm$^{-2}$ & $\zeta(r_p,z=0)$, s$^{-1}$& IG99 fit \\ \hline
10S0 & 10 & $3\times 10^{-4}$ & 1              & $2.26$             & $2.8\times 10^{-17}$ & 0.71 \\
10S2 & 10 & $3\times 10^{-4}$ & 0.01           & $2.26$             & $6.8\times 10^{-17}$ & 1.7 \\
10J0 & 10 &         $10^{-3}$ & 1              & $0.168$            & $3.2\times 10^{-16}$ & 6.3 \\
10J2 & 10 &         $10^{-3}$ & 0.01           & $0.168$            & $3.9\times 10^{-16}$ & 7.7 \\
 5S0 &  5 & $3\times 10^{-4}$ & 1              & $1.45$             & $1.8\times 10^{-16}$ & 1.0 \\
 5S2 &  5 & $3\times 10^{-4}$ & 0.01           & $1.45$             & $4.2\times 10^{-16}$ & 2.4 \\
 5J0 &  5 &         $10^{-3}$ & 1              & $0.107$            & $2.1\times 10^{-15}$ & 10  \\
 5J2 &  5 &         $10^{-3}$ & 0.01           & $0.107$            & $2.3\times 10^{-15}$ & 11  \\ \hline \\
\end{tabular}
\end{table}

\section{X-RAY IONIZATION\label{sec:transfer}}

We compute X-ray ionization rates in each model disk by sampling the
X-ray mean intensity using Monte Carlo techniques, and converting the
absorbed X-ray energies into ionization rates.  In most respects we
follow \citet[hereafter IG99]{1999ApJ...518..848I}.  In the sections
below we sketch the IG99 approach and note our points of departure.

\subsection{X-Ray Source\label{sec:source}}

Magnetic reconnection heats plasma in the young star's corona to
temperatures exceeding $kT_X = 1$~keV.  The plasma emits X-rays by
thermal bremsstrahlung, whose spectral luminosity at energy $E$ we
approximate by
\begin{equation}\label{eq:brems}
  L(E) = \frac{L_X}{kT_X} \exp \left(-\frac{E}{kT_X} \right),
\end{equation}
where $L_X$ is the total X-ray luminosity and $T_X$ is the source
temperature.  A less approximate treatment would involve quantum
electrodynamic corrections.  An
electron-velocity-distribution-averaged Gaunt factor that has been set
to unity in deriving eq.~\ref{eq:brems} would then decline by a factor
four across the energy range from 1 to 30~keV, tilting the spectrum
toward low energies \citep{1961ApJS....6..167K}.  This would yield
higher ionization rates in the disk's surface layers, and lower rates
in the interior.  However, we follow IG99 in using eq.~\ref{eq:brems}.

Also like IG99, we locate the X-ray source in a helmet streamer taking
the shape of a ring centered on the rotation axis, whose radius and
height above the disk midplane are each ten times the stellar radius.
We adopt a source temperature $T_X=5$~keV and an X-ray luminosity
$L_X=2\times 10^{30}$~erg~s$^{-1}$, the median for solar-mass stars in
the Orion Nebula Cluster \citep{2000AJ....120.1426G}.  We consider
X-rays below 1~keV to be attenuated in the star's magnetosphere and
inner wind.  We thus compute the transfer of X-rays with energies from
1 up to 30~keV.

\subsection{X-Ray Opacities}

\begin{figure}[tb!]
\centering
\includegraphics[scale=0.75]{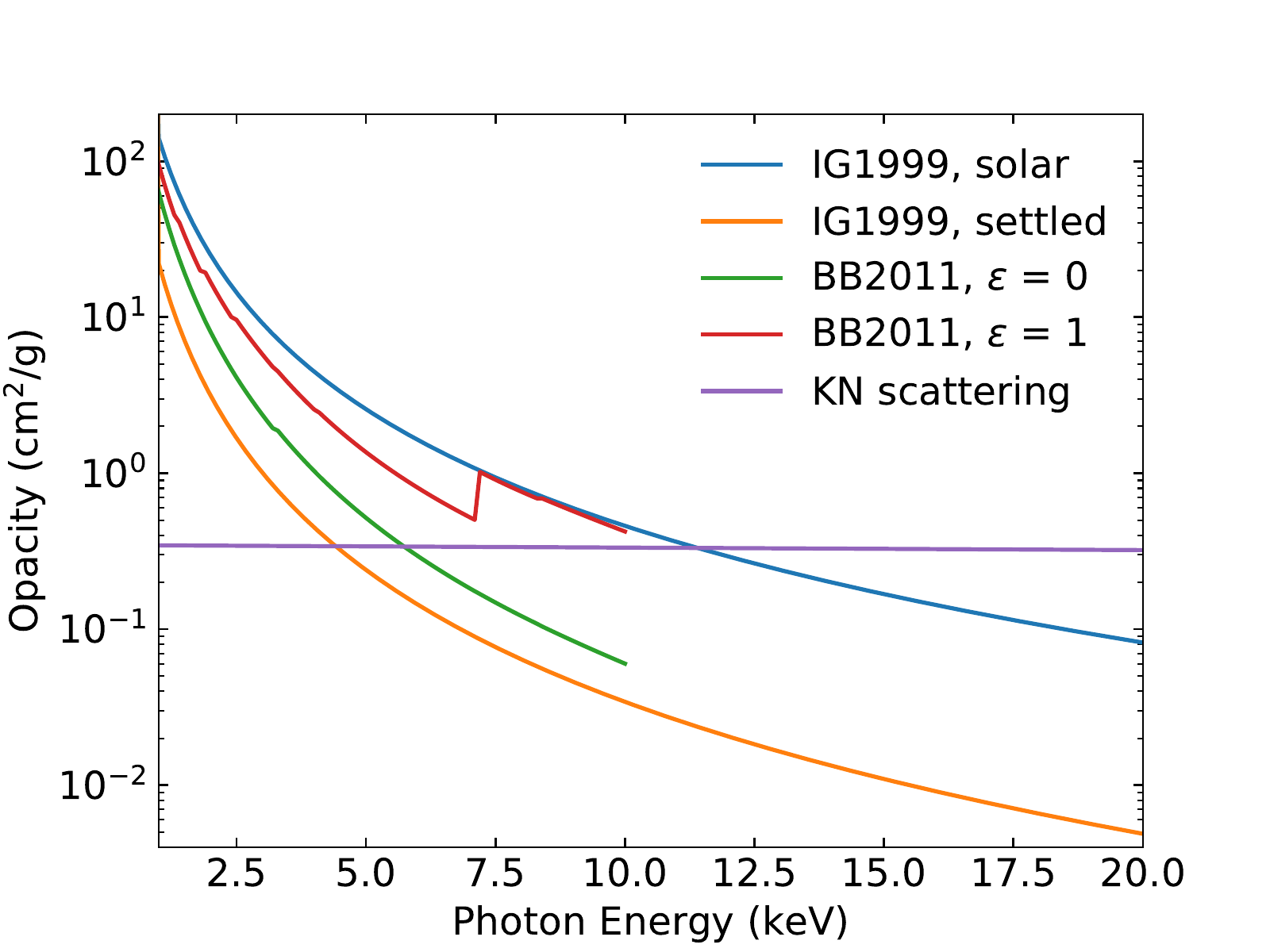}
\caption[X-ray opacities]{X-ray scattering and absorption opacities.
  The nearly-horizontal line denotes the scattering opacity derived
  from the Klein-Nishina formula.  The absorption opacities, from top
  to bottom, are (1) IG99's power-law fit assuming solar abundances,
  (2) \citet{2011ApJ...740....7B}'s more detailed fit with dust
  undepleted and (3) fully depleted, and (4) IG99's power-law fit for
  solar abundance but with heavy elements depleted onto grains which
  have been removed by settling and growth.  We use the Bethell \&
  Bergin opacities.}
\label{fig:opacities}
\end{figure}

IG99 used a simple power-law energy dependence for the X-ray
absorption cross section, and carried out their Monte Carlo
calculations assuming all heavy elements were segregated from the gas.
We adopt a more detailed fit by \citet{2011ApJ...740....7B}, who for
solar elemental abundances find a cross section of the form
\begin{equation}
  \sigma_{\text{tot}} = \sigma_{\text{gas}} + \epsilon f_b(E) \sigma_{\text{dust}},
\end{equation}
where $\epsilon$ and $f_b$ are the dust settling and grain growth
parameters, respectively, and the cross sections of the gas and dust
are each given by
\begin{equation}
  \sigma(E) = 10^{-24} \text{cm$^2$} \times (c_0 + c_1 E + c_2 E^2) E^{-3},
\end{equation}
with different fitting coefficients $c_i$ for dust and gas.  The
coefficients are piecewise linear functions of energy, to reproduce
discontinuous increases in the absorption opacity due to K-shell
photoelectric absorption by various metals.
\citet{2011ApJ...740....7B} provide fits up to 10~keV; we extrapolate
to higher energies.  Consistent with the optical and infrared
transfer, we consider both $\epsilon=1$ and $\epsilon=0.01$ to allow
for the depletion of small grains.  The dust growth parameter $f_b$ is
unity for all our calculations.  Like IG99 we take the scattering
cross section from the Klein-Nishina formula.  Cross sections are
converted to opacities using a mean gas molecular weight of 2.3,
corresponding to solar composition with the hydrogen in molecular
form.  The energy dependence of the opacities is shown in
Figure~\ref{fig:opacities}.  The scattering cross section is almost
constant across the energy range shown, falling off very slightly
toward the top end.  The absorption falls off much more steeply with
photon energy.  Thus the albedo $\omega = \sigma/(\kappa+\sigma)$,
where $\kappa$ is the absorption and $\sigma$ is the scattering
opacity, increases dramatically with energy.  The highest-energy
photons are more likely to be scattered than absorbed.  This
contributes to the harder X-rays penetrating deeper into the disk.

\subsection{X-Ray Transfer and Ionization Rate}

We sample the X-rays' mean intensities on the radiative transfer grid
using a Monte Carlo approach.  We follow each packet of X-ray photons
from its emission at the stellar source into the disk and through as
many scatterings as needed till the packet either escapes the domain
or is absorbed.  We assume that all absorbed X-ray energy is converted
into ionization.  A scattered X-ray packet's new direction is chosen
randomly from the Klein-Nishina phase function and its energy is
reduced by Compton losses.

X-ray sources with different energies produce mutually independent
intensities in the disk.  We therefore perform the Monte Carlo
procedure separately for each monochromatic source energy.  To
construct the desired source spectrum we linearly combine the
intensities produced by the monochromatic sources.  The resulting mean
X-ray intensity is converted to an ionization rate by dividing by the
average energy required to produce an ion pair, for which we follow
IG99 and adopt the value $\Delta E$ = 37~eV.

We differ from IG99 in how we compute the mean intensity.  They
average over the projected area met by each packet at each radiative
transfer grid cell.  We instead follow \citet{1999A&A...344..282L} in
integrating the photon packet's contribution all along its path.  This
yields lower Monte Carlo noise for a given number of packets,
particularly in the optically-thin disk atmosphere.

As shown in figure~\ref{fig:midplane_spectrum} top panel, the source
spectrum is adequately represented with 30 energy bins.  Thus, for
each of 30~monochromatic X-ray source energies, uniformly spaced from
1~to 30~keV, we send $10^6$~photon packets into our model disks, then
construct a weighted sum of the monochromatic results to determine
ionization rates for a source with a thermal bremsstrahlung spectrum
at a temperature of 5~keV.

\section{MAGNETIC COUPLING\label{sec:chemistry}}

\subsection{Ionization State\label{sec:ionization}}

We compute the ionization state by balancing the X-ray ionization with
a recombination reaction network including grain surface reactions and
simplified gas-phase chemistry \citep[model~4
  of][]{2006A&A...445..205I}.  This yields the equilibrium abundances
of seven charged species: electrons, a representative molecular ion
(HCO$^+$), a representative metal ion (Mg$^+$), and grains charged by
one and two electrons either side of neutral.

The magnesium has abundance $3.39 \times 10^{-7}$ atoms per hydrogen
nucleus, 1\% of the solar value, since most of the magnesium is locked
up in minerals and only a minority can react on the grain surfaces or
in the gas phase.  The fraction of magnesium in available form has
little impact on the ionization balance, since temperatures are low
enough that most magnesium atoms stay adsorbed on the grains.

The grains' abundances in the reaction network match those in the
corresponding starlight and X-ray radiative transfer calculations, in
the following approximate sense.  The reaction network treats grains
of a single radius, which we set to 0.1~$\mu$m.  When combined with a
material density of 2~g~cm$^{-3}$, this nearly matches the geometric
cross-section per unit mass in the optical opacity model.  Extending
the grain population from a single size to a power-law size
distribution yields recombination rates that depend a little
less-steeply than linearly on the total cross-section
\citep{2009ApJ...701..737B}.

\subsection{Magnetic Field Strengths}

We evaluate three scenarios for whether magnetic torques can drive the
accretion flow: (1) magneto-rotational turbulence (henceforth MRT),
(2) the Hall-shear instability (HSI), and (3) a magneto-centrifugal
wind launched from the disk surface.  The MRT sustains a tangled field
with strong azimuthal and moderate radial components, on a weaker
vertical background field.  Hall-shear instability couples the Hall
term's rotation of toroidal into radial field, together with the
orbital shear's stretching out the radial component to generate fresh
toroidal field.  The resulting fields have a dominant toroidal
component.  The magneto-centrifugal winds we consider have all three
field components comparable near the disk surface.

For each scenario to produce the accretion rates present in the model
disk, the field must reach a certain minimum strength, as follows.
Taking the equation of motion in cylindrical coordinates $(r,\phi,z)$,
reducing to the case of a near-axisymmetric, near-Keplerian disk with
time-steady mean internal flows, and averaging over the disk thickness
$2h$, yields
\begin{equation}\label{eq:mdot}
  {{\dot M}\Omega\over 2r} = {2h\over r}\left<-B_rB_\phi\right> +
  {\partial\over\partial r}\left[h\left<-B_rB_\phi\right>\right] -
  (B_zB_\phi)_s,
\end{equation}
where angle brackets denote averages through the disk, and the
subscript $s$ marks fields measured on the disk's top and bottom
surfaces \citep{2007Ap&SS.311...35W}.  The first two terms on the
right come from magnetic stresses in the disk interior, for example
due to MRT or HSI, while the final term is the wind's back-reaction on
the disk via magnetic torques.  As is common, we assume the mean
stress varies only over length scales at least comparable to the
radius, so that the second term is comparable to or smaller than the
first, and can be neglected.

In the MRT scenario, angular momentum conservation thus links the mass
flow rate $\dot M$ carried by the disk to the magnetic accretion
stress by
\begin{equation}\label{eq:mdot1}
  \left<-B_rB_\phi\right>={{\dot M}\Omega\over 4h}.
\end{equation}
The stress from the field's radial and azimuthal components is about
one-quarter the squared magnitude of the magnetic field, which in turn
is about 20~times the mean squared vertical field, based on direct
numerical calculations \citep{1995ApJ...440..742H,
  2004ApJ...605..321S}.  We can thus solve eq.~\ref{eq:mdot1} for the
RMS vertical field.

Eq.~\ref{eq:mdot1} also gives the stress in the HSI scenario.  To find
the mean vertical field strength, we observe that the toroidal
component is about 50 and 200 times the radial and vertical components
in \citet[run 1-OHA-5]{2014A&A...566A..56L}.

In the wind scenario, eq.~\ref{eq:mdot} becomes
\begin{equation}\label{eq:mdot3}
  (-B_zB_\phi)_s={{\dot M}\Omega\over 2r},
\end{equation}
where the $s$ subscript indicates values measured at the disk surface.
We estimate the vertical field using the fact that magneto-centrifugal
wind solutions typically have the three field components roughly equal
at the disk surface.  The surface connects the interior, with its
vertical magnetic field, to the wind, where the field is angled away
from the rotation axis.

\subsection{Coupling Criteria\label{sec:couplingcriteria}}

With the field strength in hand from eq.~\ref{eq:mdot1}
or~\ref{eq:mdot3}, and the charged species' abundances from
sec.~\ref{sec:ionization}, we can find the magnetic diffusivities and
determine whether the field in fact couples to the gas well enough to
drive the accretion flow.

We compute the diffusivities $\eta_O$, $\eta_H$ and $\eta_A$ that are
the coefficients of the Ohmic, Hall and ambipolar terms in the
induction equation, including the contributions from all charged
species, following \citet{2007Ap&SS.311...35W} eqs.~21-31.  The
diffusivities depend on the field strength, which governs whether each
charged species mostly gyrates around the field lines under the
Lorentz force, or mostly random walks by colliding with neutrals.

Each of the three magnetic scenarios works only when the diffusivities
and field strengths meet a set of requirements.  For MRT, disturbances
with the linear magneto-rotational instability's fastest-growing
wavelength must diffuse away more slowly than they grow.  The
wavelength $v_{Az}/\Omega$ depends on the Alfven speed $v_{Az}$ along
the vertical magnetic field, the growth rate is close to $\Omega$, and
the relevant diffusivity is the sum of the Ohmic and ambipolar values,
called the Pedersen diffusivity $\eta_P = \eta_O + \eta_A$.  Thus MRT
requires
\begin{equation}\label{eq:etaPthreshold}
  \eta_P < v_{Az}^2/\Omega
\end{equation}
\citep{2001ApJ...561L.179S, 2007ApJ...659..729T, 2015MNRAS.451.1104K}.
In addition the wavelength must fit within the disk thickness,
corresponding to a vertical magnetic field with pressure less than
that of the gas by a factor
\begin{equation}\label{eq:betazthreshold}
\beta_z > 8\pi^2
\end{equation}
\citep{2011ApJ...742...65O}.  At the same time, the Hall term must be
small enough to not much modify the character of the turbulence.
\citet{2014A&A...566A..56L} find that the Hall term dominates if the
Hall length exceeds 20\% of the gas scale height.  The Hall length is
$|\eta_H|/v_A$, thus standard MRT requires approximately
\begin{equation}\label{eq:lHthreshold}
  |\eta_H| < \beta_z^{1/2} v_{Az}^2/\Omega.
\end{equation}

For the Hall-shear instability in contrast, a strong Hall term is
required: the Hall diffusivity must exceed the right-hand side of
eq.~\ref{eq:lHthreshold}.  Furthermore a large Hall term appears able
to drive instability even in the face of significant Ohmic and
ambipolar diffusion \citep{2014A&A...566A..56L}.  We therefore place
no upper limit on the Pedersen diffusivity under which the Hall-shear
process can operate.  If the fastest-growing wavelength were the
determining factor, a too-strong magnetic field would again be
disqualifying, and eq.~\ref{eq:betazthreshold} would also be a
requirement for Hall-shear instability \citep{2015MNRAS.451.1104K}.
However we note that the spectrum of unstable linear modes can extend
to wavelengths shorter than the ideal-MHD cutoff if the Hall term is
strong \citep{2002ApJ...570..314S}, and that the Hall-shear
instability operates with a midplane plasma beta near unity in
\citet{2014A&A...566A..56L} run 1-OHA-5.  For these reasons we do not
limit the field strengths at which Hall-shear instability is allowed.

Finally, to launch a magneto-centrifugal wind, the disk must be able
to sustain vertical gradients in the field's horizontal components.
That is, the fields may not diffuse through the disk thickness in less
than one orbit.  This boils down to
\begin{equation}\label{eq:windthreshold}
  \eta_{\rm max} < c_s^2/\Omega,
\end{equation}
where $\eta_{\rm max}$ is whichever is larger, $\eta_P$ or $\eta_H$
\citep{2015MNRAS.451.1104K}.

To summarize, each magnetic scenario and mass flow rate imply a field
strength, which together with the charged species' populations yields
the diffusivities.  The MRT scenario is viable if the field and
diffusivities satisfy eqs.~\ref{eq:etaPthreshold},
\ref{eq:betazthreshold} and~\ref{eq:lHthreshold}.  The Hall-shear
scenario is viable if the negation of eq.~\ref{eq:lHthreshold} holds.
The magnetized wind is viable given eq.~\ref{eq:windthreshold}.  To
express these requirements compactly below, we define $\eta_{v_{Az}}
\equiv v_{Az}^2/\Omega$ and $\eta_{c_s} \equiv c_s^2/\Omega$.

\section{RESULTS\label{sec:results}}

\subsection{X-ray Spectra}

\begin{figure}[tb!]
\centering
\includegraphics[scale=0.75]{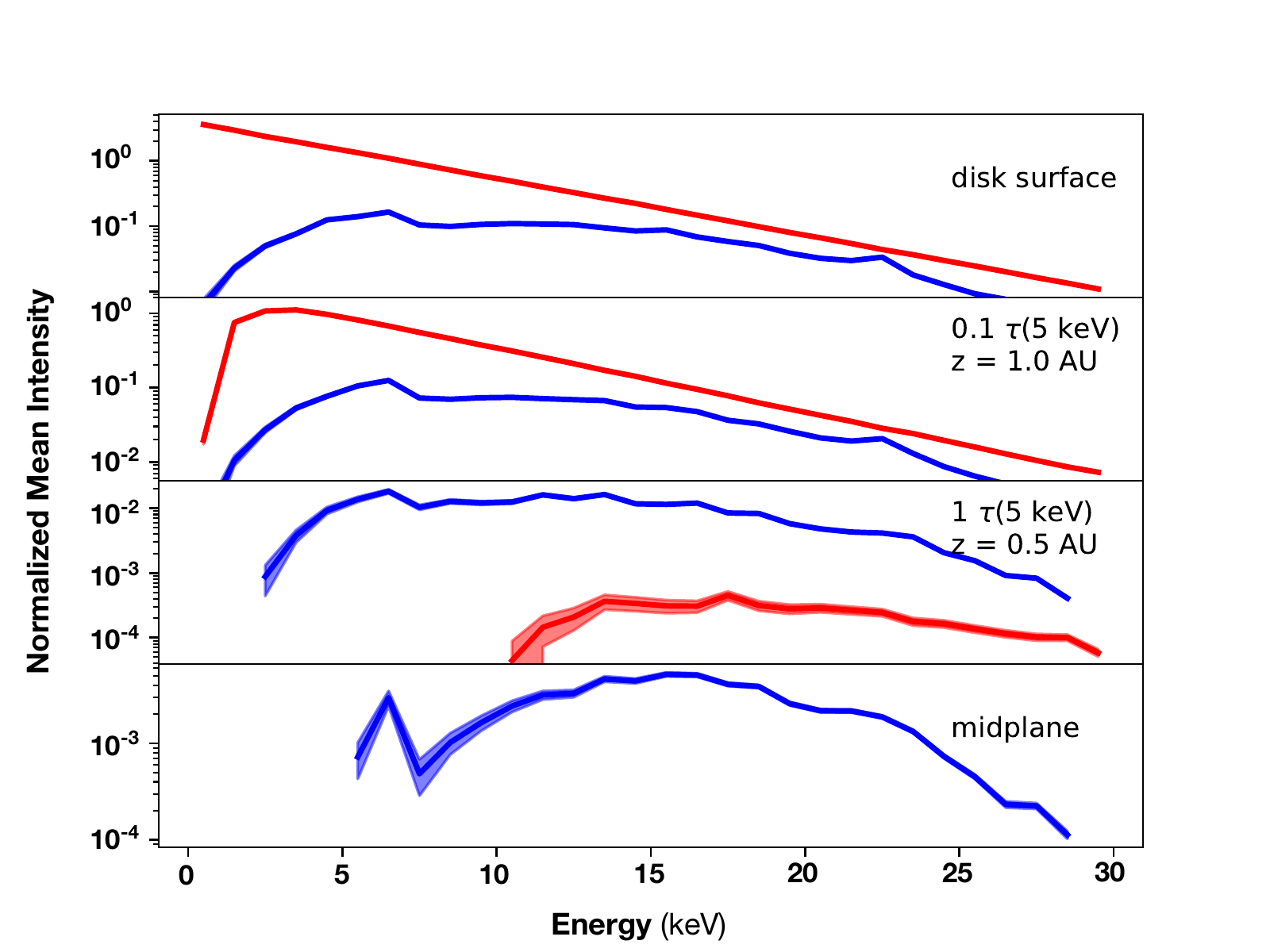}
\caption{X-ray spectra in model 10S0 at four locations above the
  planet.  The top panel is at the disk surface, the next two are 1.0
  and 0.5~au above the planet, and the bottom panel is at the
  midplane, on the planet's orbit.  Each is labeled with the vertical
  optical depth at 5~keV.  Red curves show the photons received
  directly from the source, and blue curves the scattered photons.
  Each curve is surrounded by lighter shading marking the $1/\sqrt{N}$
  uncertainty in the Monte Carlo results.  All are normalized to the
  mean intensity (erg/cm$^2$/sr/s/keV) at the disk surface at 5~keV.
  Direct photons are thoroughly absorbed for vertical optical depths
  unity and greater.  Only X-rays with energies over 5~keV reach the
  midplane in significant numbers, and all these have been scattered.}
\label{fig:midplane_spectrum}
\end{figure}

In all cases, the X-rays reach the midplane with a much lower
intensity than in the disk atmosphere, and a spectrum favoring higher
energies.  In figure~\ref{fig:midplane_spectrum} are the spectra
observed at four heights above the planet in model 10S0.  The
lower-energy direct X-rays are absorbed high in the atmosphere, while
higher-energy photons' intensity declines with depth, as Compton
down-scattering converts them to lower energies where they are more
easily absorbed.  Few or no photons reach the planet with energies
below 10~keV, while the intensity at energies above 10~keV is around
0.1\% of that expected in the absence of the intervening disk
material.  Almost all photons reaching the midplane were emitted from
the source with energies above 18~keV.  The 8~keV dip in the spectrum
at intermediate heights comes from the iron absorption edge near
7~keV, visible in figure~\ref{fig:opacities}.

\begin{figure*}[tb!]
\centering
\includegraphics[scale=0.5]{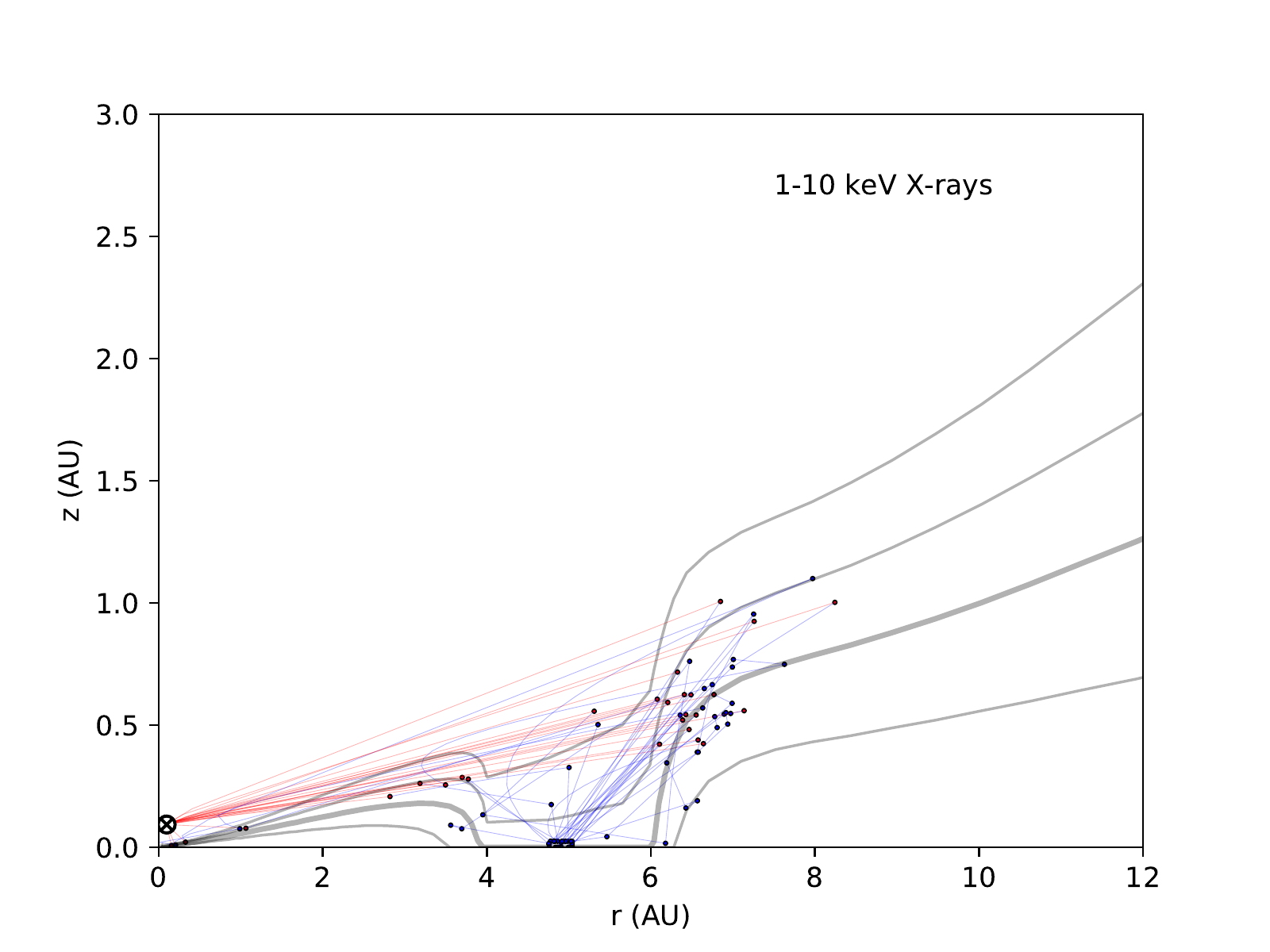}
\includegraphics[scale=0.5]{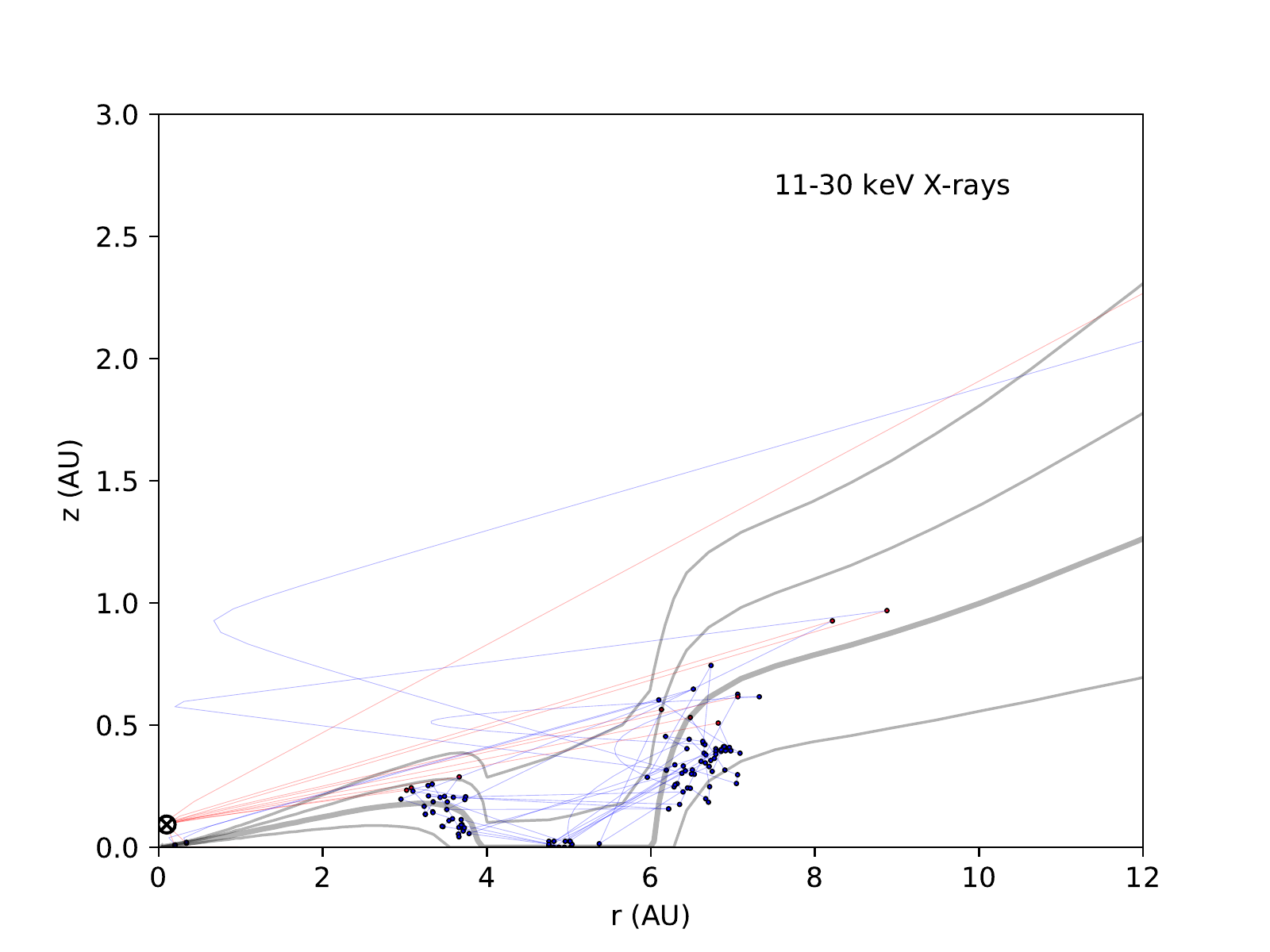}
\caption{Paths in the $(r,z)$ plane of photon packets chosen randomly
  from among those reaching the planet's vicinity in model 5J0, which
  has a Jupiter-mass planet at 5~au in a dusty disk.  The X-ray source
  is marked by the cross in a circle near $(0.1,0.1)$~au.  Each path
  is red before the first scattering, and blue thereafter.  Filled
  circles mark scattering points.  The thick grey line shows the
  surface of unit vertical optical depth to 5-keV photons, while thin
  grey lines denote optical depths spaced by factors of ten.  The
  first panel shows photons with energies up to 10~keV, for which the
  single-scattering albedo is below 50\%.  Most photons reaching the
  planet are scattered just once off the gap rim.  The second panel is
  for higher-energy, higher-albedo photons, which often scatter
  repeatedly off the gap walls.  Some scatter first in the disk
  surface layers interior to the gap, and a few pass near the rotation
  axis as they cross from one side of the disk to the other.  The
  number of packets in each panel is in proportion to the energy
  band's contribution to ionization near the planet.  }
\label{fig:gap_scattering}
\end{figure*}

The cleaner gaps opened by the Jupiters let more photons with energies
below 10~keV reach the planet's vicinity
(figure~\ref{fig:gap_scattering}).  Still most of the ionization comes
from photons emitted with energies above 10~keV, many of which scatter
more than once because their single-scattering albedos exceed
one-half.

\subsection{Ionization Rates}

\begin{figure}[tb!]
\centering
\includegraphics[scale=0.5]{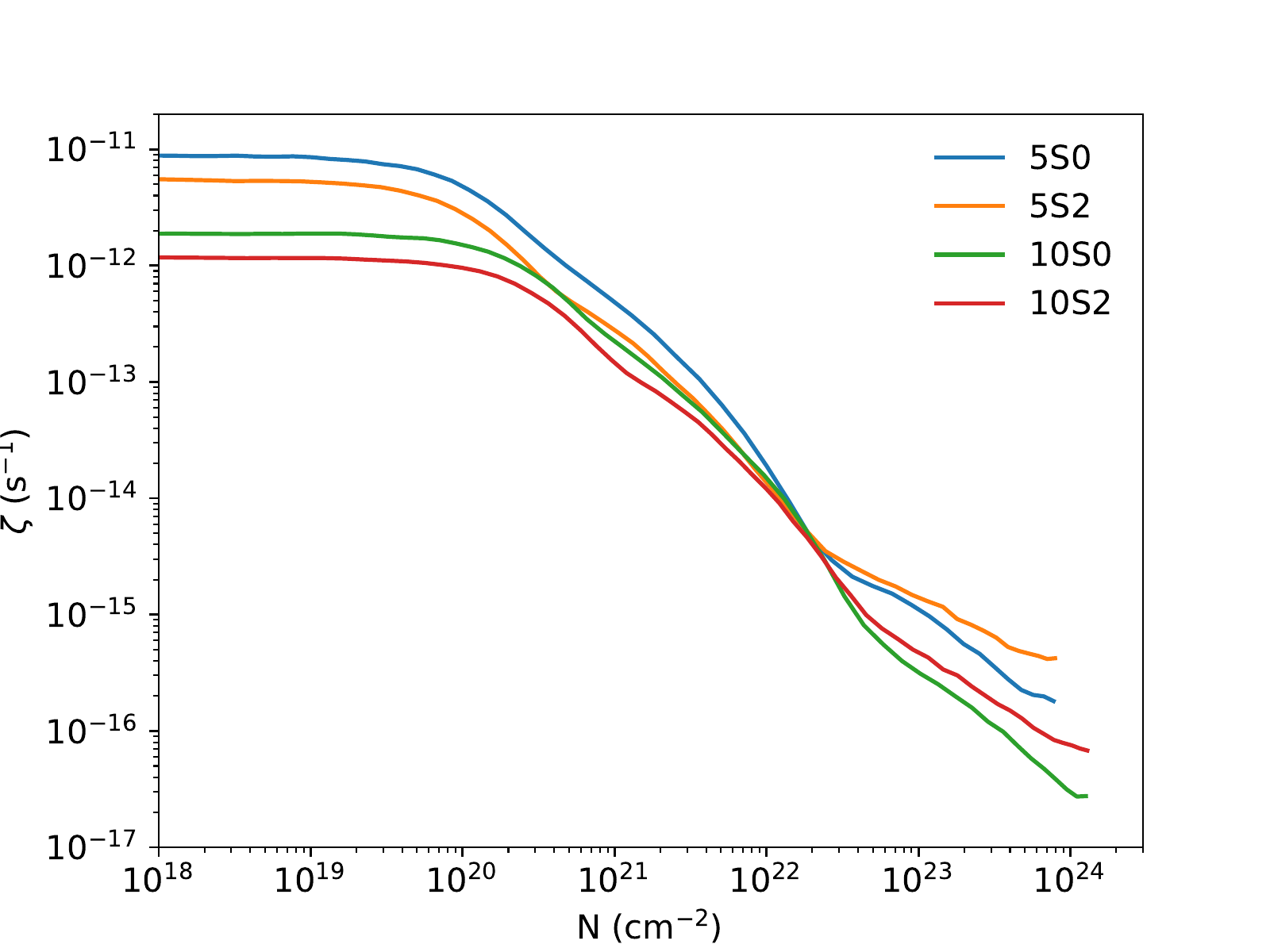}
\includegraphics[scale=0.5]{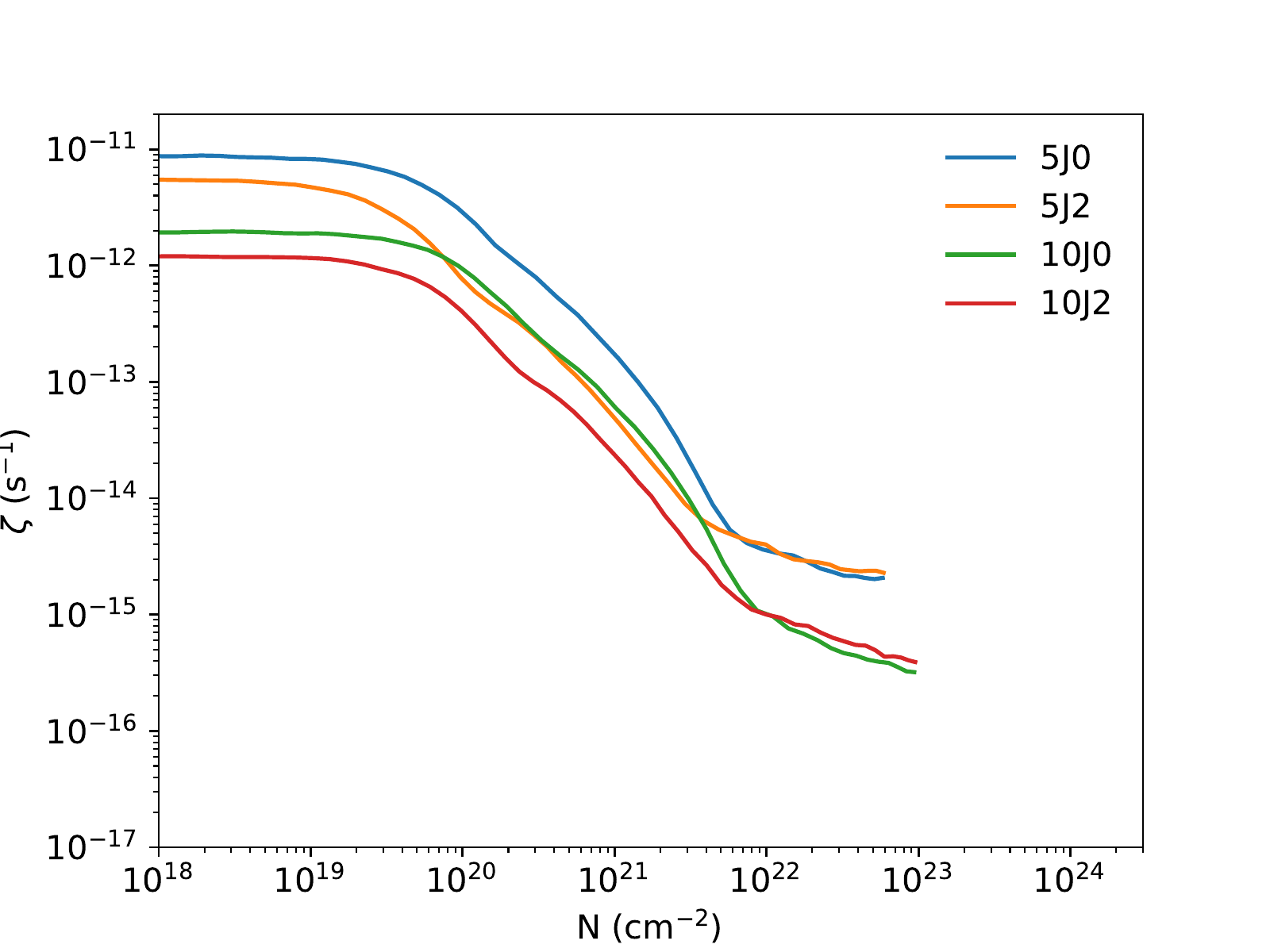}
\caption{Ionization rates vs.\ vertical column of hydrogen nuclei,
  above planets with the mass of Saturn (left) and Jupiter (right).
  The X-rays ionize faster than $10^{-16}$~s$^{-1}$ in all models
  except those with the Saturn at 10~au.}
\label{fig:ion_rates}
\end{figure}

The ionization rates for all models are plotted in
figure~\ref{fig:ion_rates} versus the column perpendicular to the
midplane.  In all cases the rates are column-independent in the
optically-thin upper layers, where they are fixed by the flux of the
numerous low energy X-rays.  At greater columns, the disk is
optically-thick to the softer X-rays, precipitating a sharp decline in
the ionization rate.  A scattering shoulder appears at columns greater
than $N_H = 10^{22}$~cm$^{-2}$, or vertical 5-keV optical depths
greater than~0.05, where the contribution from scattered harder X-rays
extends ionization into the disk interior.  The ionization rate
asymptotes at the highest columns in the Jupiter cases, because the
total X-ray optical depth is low, so all points near the midplane see
the remaining scattered X-rays.

At greater distances from the star, X-rays are less important because
of the inverse-square falloff in their flux.  It is thus worth
considering whether the X-rays are competitive with the interstellar
cosmic rays.  These are attenuated by the protostar's wind
\citep{2013apj...772....5i}, which however was likely funneled by the
disk and its wind into a bipolar configuration.  While the disk at
10~au could receive some cosmic rays focused by the disk wind or
entering near the equatorial plane, molecular abundances observed at
tens of au in at least one disk are well fit by a model without cosmic
rays \citep{2015ApJ...799..204C}.  Thus the cosmic ray ionization rate
inside 10~au is likely well below the interstellar value of about
$10^{-17}$~s$^{-1}$.  Thermal ionization and radionuclide decay
ionization \citep{2009ApJ...690...69U} are orders of magnitude weaker
still at 5 and 10~au.

X-ray ionization rates at the planet exceed $10^{-16}$~s$^{-1}$ in all
the models we consider except those with a Saturn-mass planet at
10~au.  Specifically, the X-ray ionization rates listed in the
next-to-last column of table~\ref{tab:models} are around $3\times
10^{-17}$ and $3\times 10^{-16}$~s$^{-1}$ for a Saturn- and
Jupiter-mass planet, respectively, at 10~au.  Moving the planet inward
to 5~au increases the ionization rate about one order of magnitude.
Reducing the dust abundance increases the X-ray albedo and so the
ionization rate near the planet.

The final column in table~\ref{tab:models} shows the ratio of the
ionization rate near the planet to that at the same radius and mass
column in the fit to the IG99 results \citep{2008ApJ...679L.131T} that
was considered by \citet{2015MNRAS.451.1104K}.  For almost all the
models, the ratio exceeds unity.  Three factors contribute.  First,
where IG99 ended their calculations at 20~keV, we include photons
emitted with energies up to 30~keV, raising ionization rates by about
10\% at the planet in the case shown in
figure~\ref{fig:midplane_spectrum}.  The X-ray source spectrum's
steepness means that further increases in the upper energy limit would
yield at most percent-level rises in the ionization rate.  Second,
IG99 assumed dust growth and settling removed all elements heavier
than helium, while we include these elements
(figure~\ref{fig:opacities}), raising by a factor of a few the
fraction of X-rays absorbed per unit column
\citep{2013MNRAS.436.3446E}, in particular among those photons that
survive to reach the planet.  Third, the gap rims scatter photons
towards the planet (figure~\ref{fig:gap_scattering}).  Only the last
of these three factors depends on the planet's mass, so the greater
enhancement over IG99 in our models with Jupiters indicates the
importance of scattering from the rims.

\subsection{Magnetic Coupling}

Inserting the ionization rates in the reaction network and evolving to
chemical equilibrium yields the ionization states.  For the dusty
cases with Saturn-mass planets, the most abundant positive and
negative species at the planet's location are molecular ions and
grains with one extra electron, respectively.  In all other cases, the
most abundant are molecular ions and free electrons.  However, in the
hottest case, with the Jupiter in the dust-depleted disk, enough metal
atoms are desorbed into the gas that the metal ions are almost as
abundant as the molecular ions.

The charged species' movements, and thus the diffusivities, depend on
the magnetic field strength as discussed in
sec.~\ref{sec:couplingcriteria}.  In addition, the coupling criteria
eqs.~\ref{eq:etaPthreshold} to~\ref{eq:lHthreshold} are explicit
functions of the field's vertical component.  Thus in
table~\ref{tab:bb} we list the total and vertical field strengths just
inside and outside the planet's orbit, obtained using
eqs.~\ref{eq:mdot1} and~\ref{eq:mdot3} from the corresponding mass
flow rates of $5\times 10^{-9}$ and $3.3\times
10^{-8}$~$M_\odot$~yr$^{-1}$ respectively.  The field strengths are
listed to two significant figures.

\newcommand{\mdoteight}{$10^{-8} M_\odot$~yr$^{-1}$}
\begin{table}[tb!]
\centering
\caption{Magnetic Field Strengths (mG) Yielding Disk Models' Mass Flow Rates\label{tab:bb}}
\begin{tabular}{rl rr rl rr} \hline
Model & ${\dot M}$ & \multicolumn{2}{c}{MRT} & \multicolumn{2}{c}{HSI} & \multicolumn{2}{c}{MCW} \\
      & \mdoteight & $B$      & $B_z$        & $B$      & $B_z$        & $B_s$      & $B_z$    \\ \hline
10S0  & 0.5        & 17       & 3.9          & 62       & 0.31         & 4.5        & 2.6      \\
      & 3.3        & 45       & 10           & 160      & 0.79         & 11         & 6.6      \\
10S2  & 0.5        & 16       & 3.6          & 58       & 0.28         & 4.5        & 2.6      \\
      & 3.3        & 41       & 9.2          & 150      & 0.73         & 11         & 6.6      \\
10J0  & 0.5        & 16       & 3.7          & 58       & 0.29         & 4.5        & 2.6      \\
      & 3.3        & 42       & 9.4          & 150      & 0.75         & 11         & 6.6      \\
10J2  & 0.5        & 15       & 3.4          & 54       & 0.27         & 4.5        & 2.6      \\
      & 3.3        & 40       & 8.8          & 140      & 0.7          & 11         & 6.6      \\
 5S0  & 0.5        & 46       & 10           & 160      & 0.82         & 11         & 6.1      \\
      & 3.3        & 120      & 27           & 420      & 2.1          & 27         & 16       \\
 5S2  & 0.5        & 42       & 9.5          & 150      & 0.75         & 11         & 6.1      \\
      & 3.3        & 110      & 24           & 390      & 1.9          & 27         & 16       \\
 5J0  & 0.5        & 44       & 9.7          & 150      & 0.77         & 11         & 6.1      \\
      & 3.3        & 110      & 25           & 400      & 2            & 27         & 16       \\
 5J2  & 0.5        & 41       & 9.2          & 150      & 0.73         & 11         & 6.1      \\
      & 3.3        & 110      & 24           & 370      & 1.9          & 27         & 16       \\ \hline \\
\end{tabular}
\end{table}

We combine the charged particle populations with the field strengths
in each of the three magnetic scenarios, to obtain the Ohmic, Hall,
and ambipolar diffusivities.  The coupling criteria are then fully
specified and we can evaluate whether each scenario is viable, in the
sense that its field strength yields diffusivities that permit the
scenario to occur.  The results for locations just inside and outside
the planet's orbit appear in table~\ref{tab:viability}.  Checks mark
scenarios that meet the conditions according to the dimensionless
numbers in columns~6 to~9.  These in turn come from
eqs.~\ref{eq:etaPthreshold} to~\ref{eq:windthreshold} with magnetic
fields from the corresponding scenarios.  The dimensionless numbers
are rounded to one significant figure.

In all cases, the MRT scenario yields a large ambipolar contribution
to the Pedersen diffusivity, which is strong enough to invalidate the
assumption that magneto-rotational turbulence is present.  In some
cases, the MRT scenario is also ruled out because it requires fields
that are too strong for the magneto-rotational wavelength to fit into
the disk thickness.  In no case is the MRT prevented solely by the
Hall diffusivity; thus we do not list the MRT scenario's Hall number
in table~\ref{tab:viability}.  In contrast, the HSI scenario implies
magnetic fields such that the Hall length is great enough for the
Hall-shear instability to operate in all except the two dust-depleted
Jupiter cases, where there is a difficulty outside the planet's orbit.
In many of the dust-depleted cases, which tend to be better-ionized,
the maximum diffusivity is low enough for the gas near the gap's
midplane to act as the base for a magneto-centrifugal wind.

\begin{table}[tb!]
\centering
\caption{Magnetic Scenarios' Viability Just Inside and Outside the
  Planet's Orbit\label{tab:viability}}
\begin{tabular}{r l c c c l l l l} \hline
Model & ${\dot M}/10^{-8}$ & MRT & HSI & Wind &MRT                   &MRT      &HSI                                  &Wind                   \\
      & $M_\odot$~yr$^{-1}$ &     &     &      &$\eta_{v_{Az}}/\eta_P$&$\beta_z$&$\beta_z^{1/2}\eta_{v_{Az}}/|\eta_H|$&$\eta_{c_s}/\eta_{\rm max}$ \\ \hline
10S0  & 0.5               & ---        & \checkmark & ---        & 6e$-$4 & 8e$+$2 & 7e$-$4 & 3e$-$1 \\
      & 3.3               & ---        & \checkmark & ---        & 4e$-$3 & 1e$+$2 & 7e$-$4 & 2e$-$1 \\
10S2  & 0.5               & ---        & \checkmark & \checkmark & 4e$-$2 & 1e$+$3 & 2e$-$2 & 5e$+$1 \\
      & 3.3               & ---        & \checkmark & \checkmark & 4e$-$2 & 2e$+$2 & 2e$-$2 & 2e$+$1 \\
10J0  & 0.5               & ---        & \checkmark & \checkmark & 1e$-$2 & 8e$+$1 & 1e$-$3 & 5e$+$0 \\
      & 3.3               & ---        & \checkmark & ---        & 2e$-$2 & 1e$+$1 & 9e$-$4 & 8e$-$1 \\
10J2  & 0.5               & ---        & \checkmark & \checkmark & 3e$-$2 & 1e$+$2 & 6e$-$2 & 2e$+$1 \\
      & 3.3               & ---        & ---        & \checkmark & 3e$-$2 & 2e$+$1 & 1e$+$0 & 3e$+$0 \\
 5S0  & 0.5               & ---        & \checkmark & ---        & 2e$-$3 & 3e$+$2 & 9e$-$4 & 4e$-$1 \\
      & 3.3               & ---        & \checkmark & ---        & 1e$-$2 & 4e$+$1 & 9e$-$4 & 3e$-$1 \\
 5S2  & 0.5               & ---        & \checkmark & \checkmark & 5e$-$2 & 4e$+$2 & 2e$-$2 & 3e$+$1 \\
      & 3.3               & ---        & \checkmark & \checkmark & 5e$-$2 & 5e$+$1 & 2e$-$2 & 1e$+$1 \\
 5J0  & 0.5               & ---        & \checkmark & \checkmark & 2e$-$2 & 2e$+$1 & 1e$-$3 & 4e$+$0 \\
      & 3.3               & ---        & \checkmark & ---        & 2e$-$2 & 4e$+$0 & 6e$-$4 & 6e$-$1 \\
 5J2  & 0.5               & ---        & \checkmark & \checkmark & 4e$-$2 & 3e$+$1 & 8e$-$2 & 1e$+$1 \\
      & 3.3               & ---        & ---        & \checkmark & 4e$-$2 & 4e$+$0 & 1e$+$0 & 2e$+$0 \\ \hline \\
\end{tabular}
\end{table}

\begin{figure}[htb!]
\centering
\includegraphics[scale=0.39]{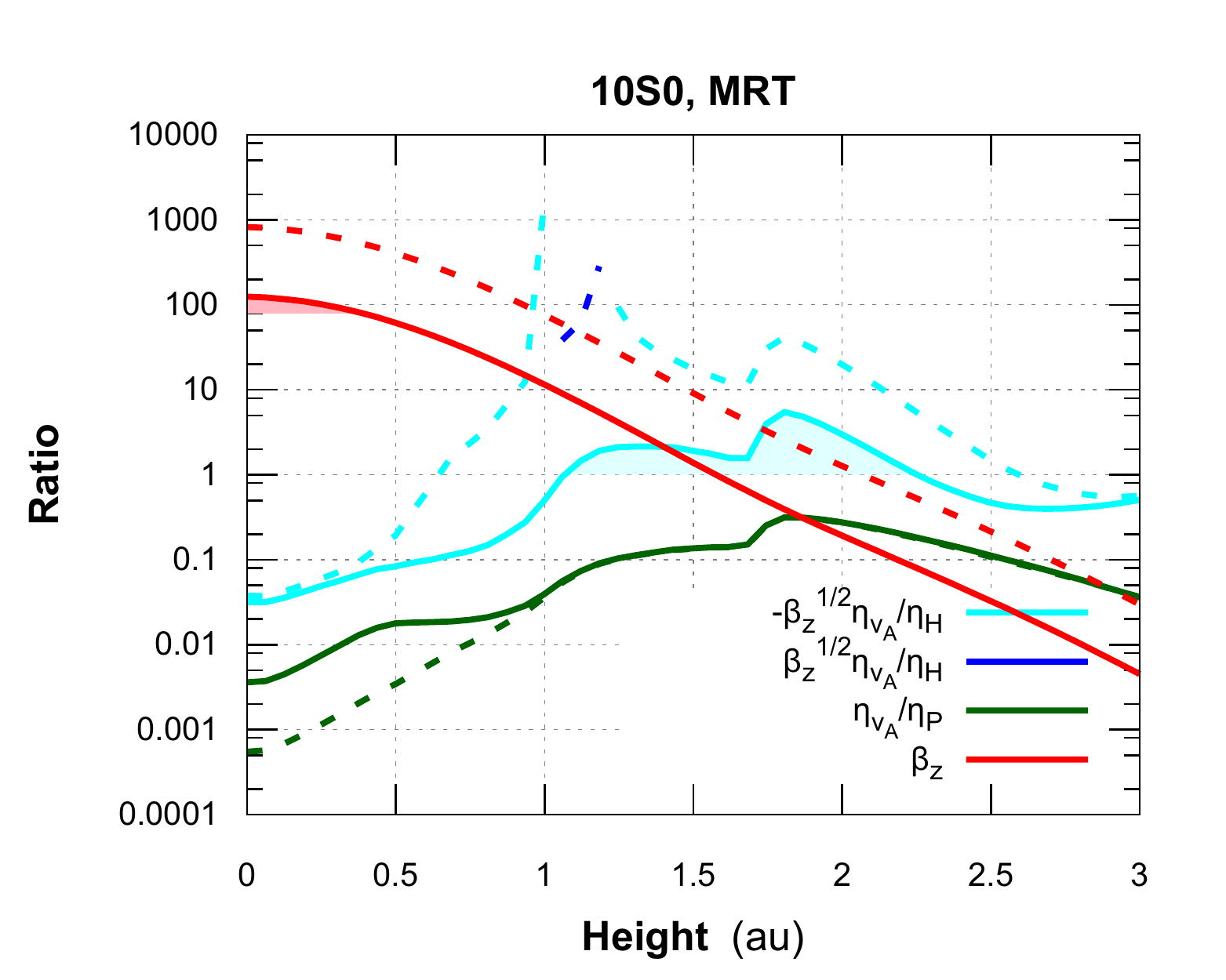}
\includegraphics[scale=0.39]{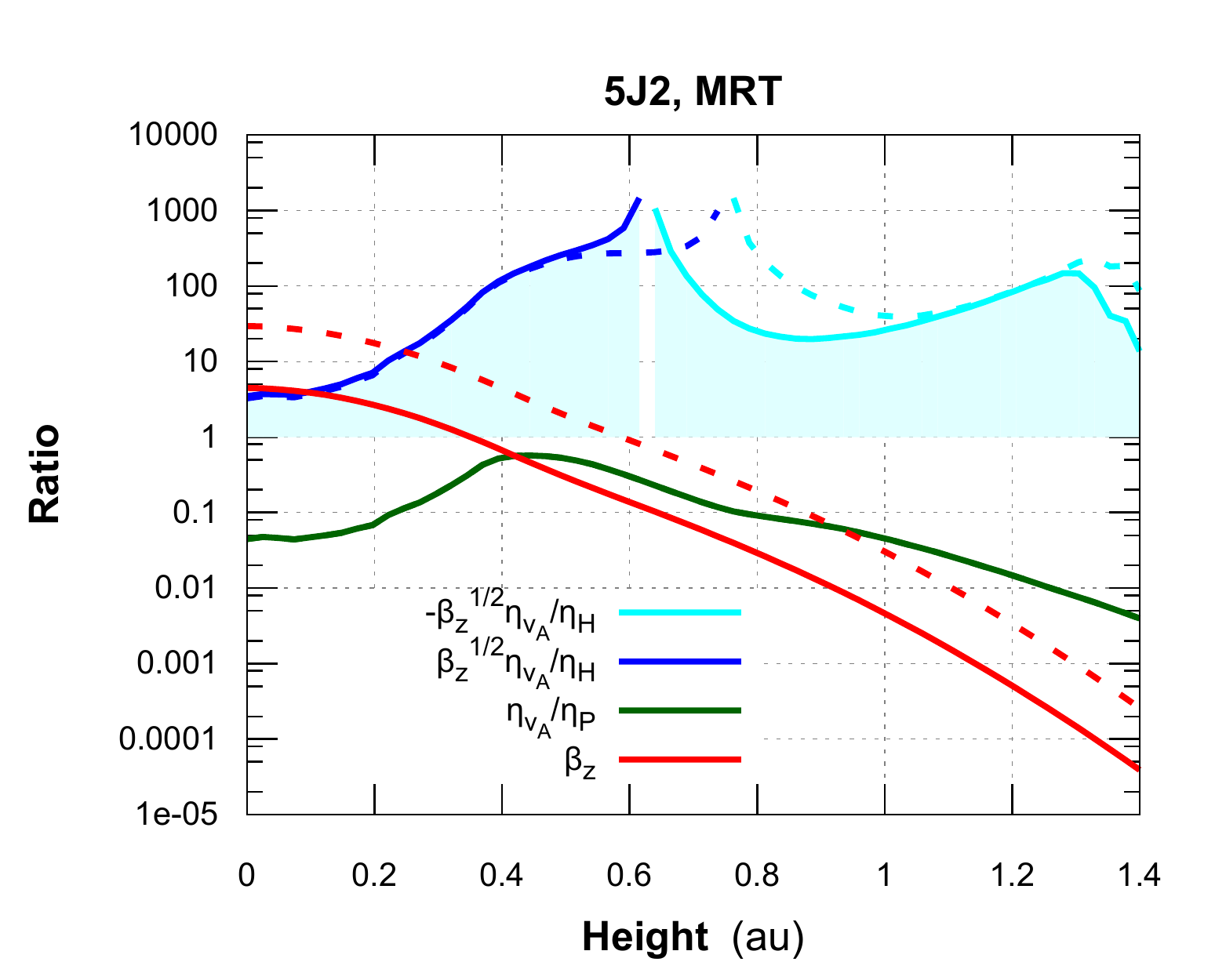} \\
\includegraphics[scale=0.39]{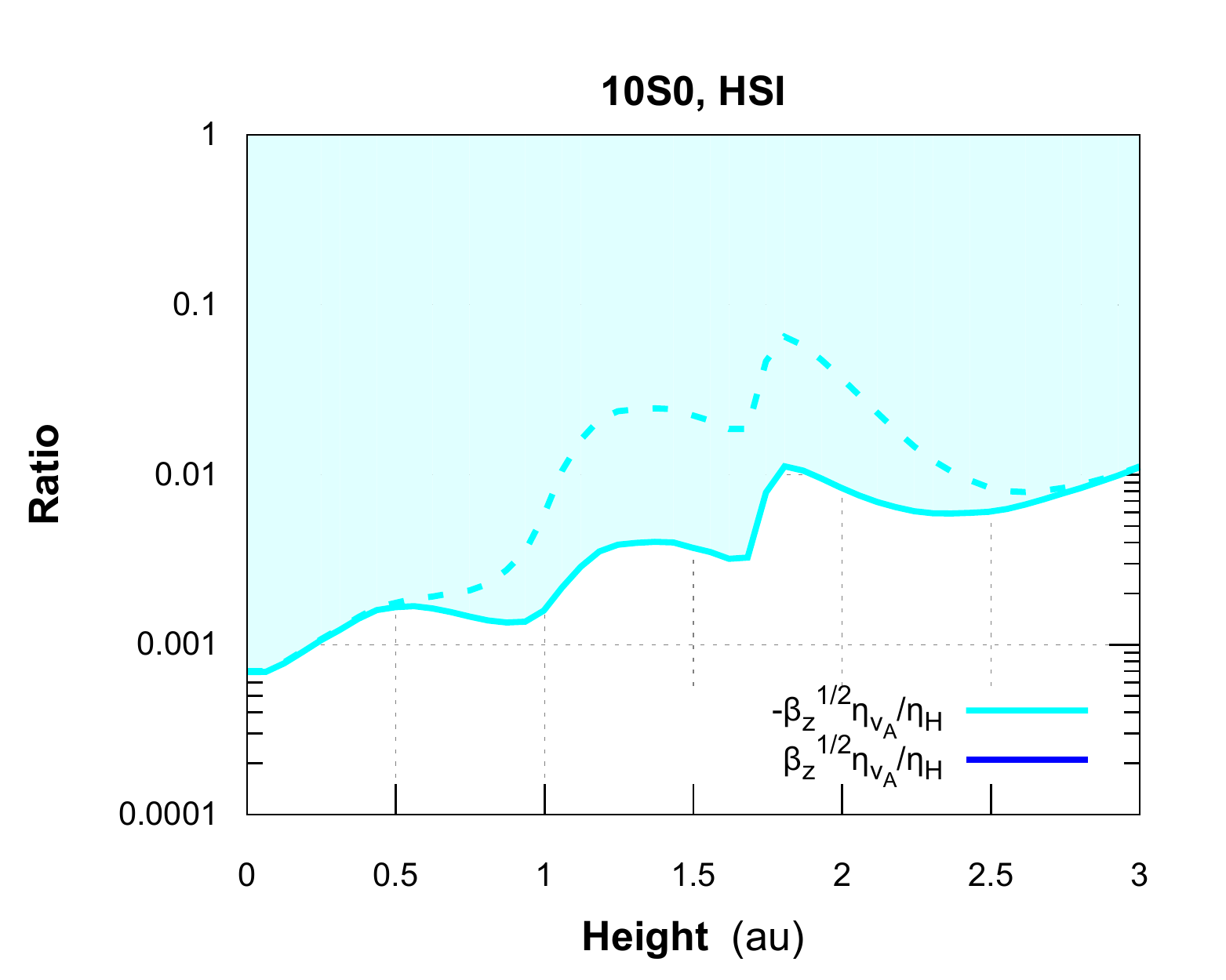}
\includegraphics[scale=0.39]{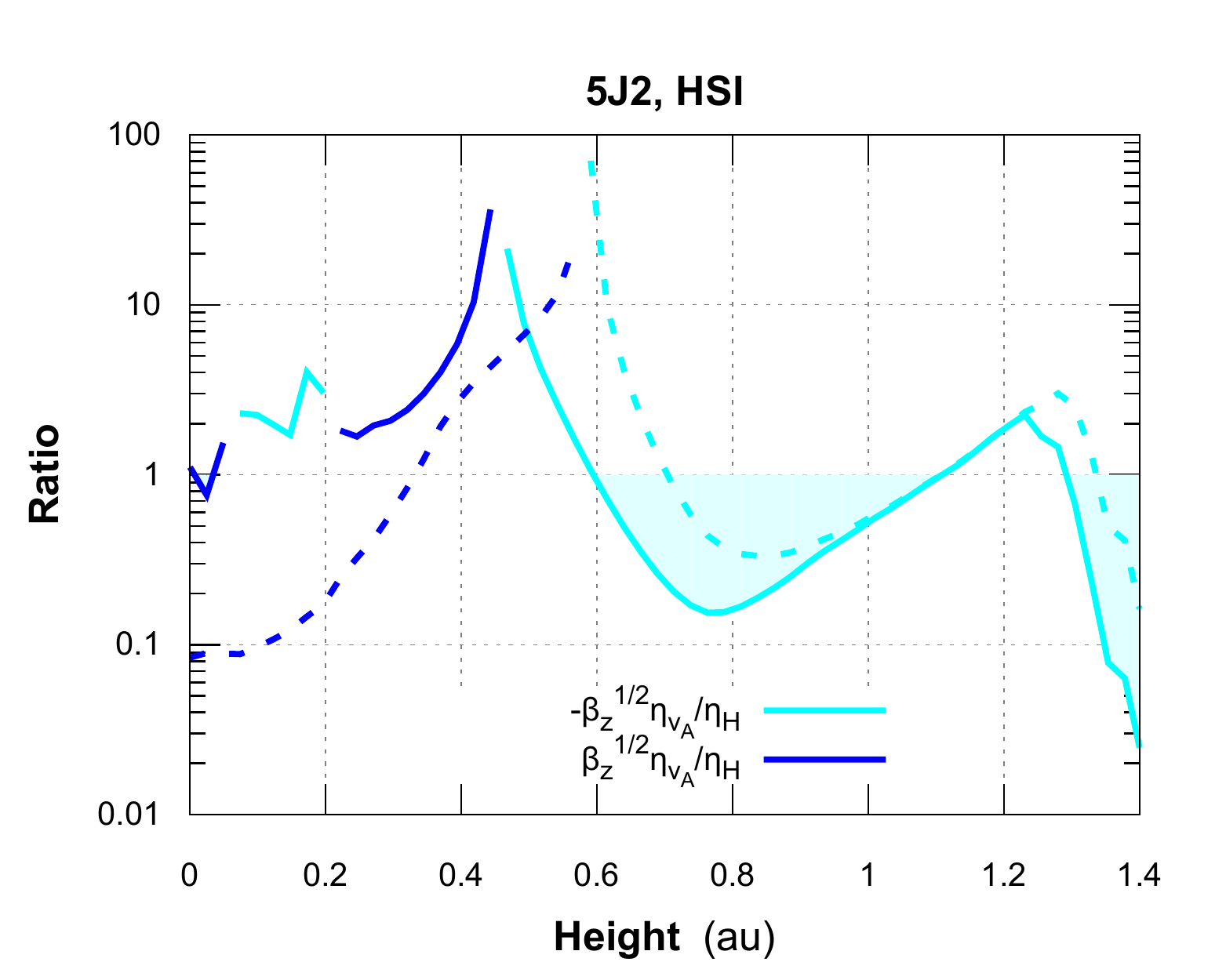} \\
\includegraphics[scale=0.39]{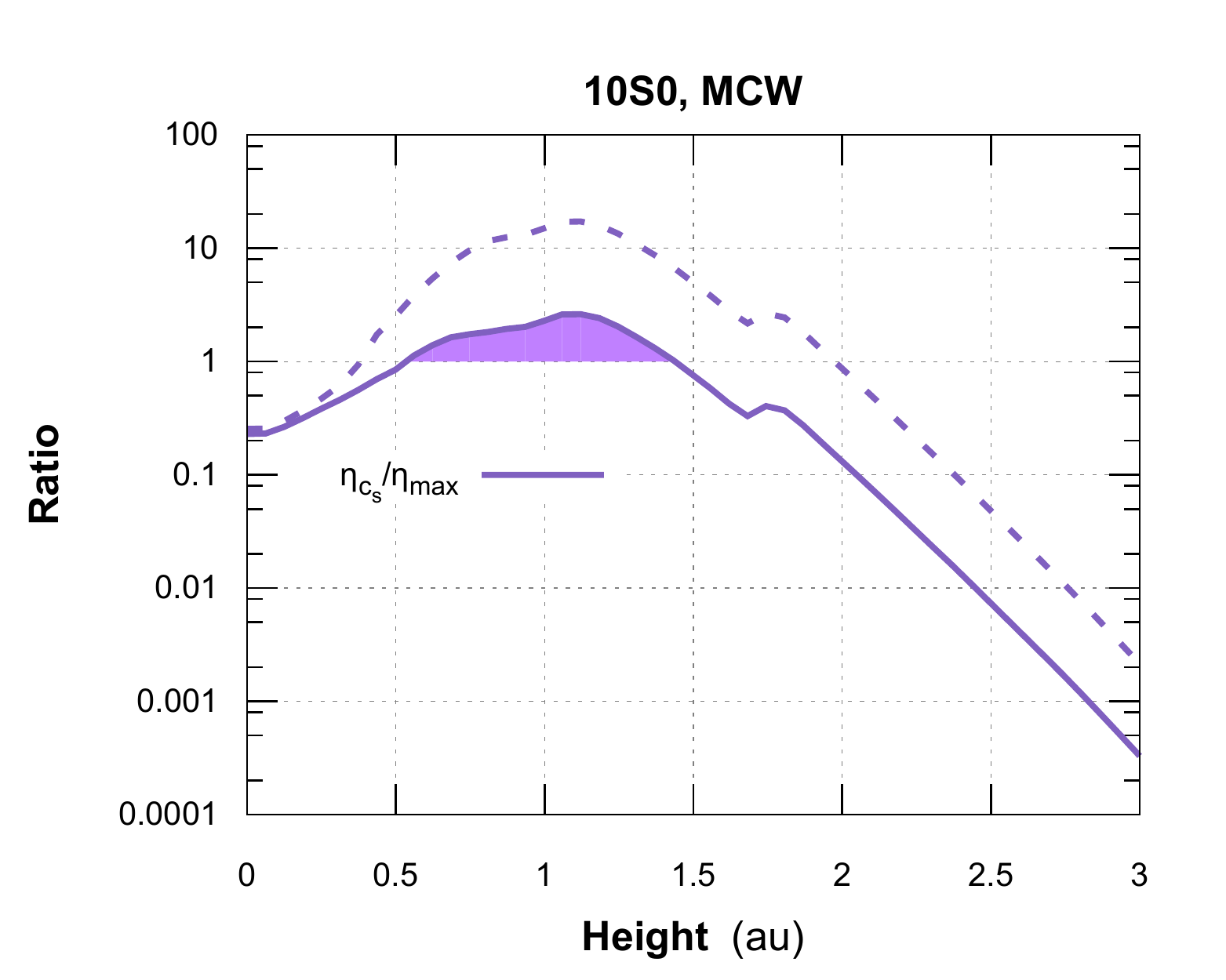}
\includegraphics[scale=0.39]{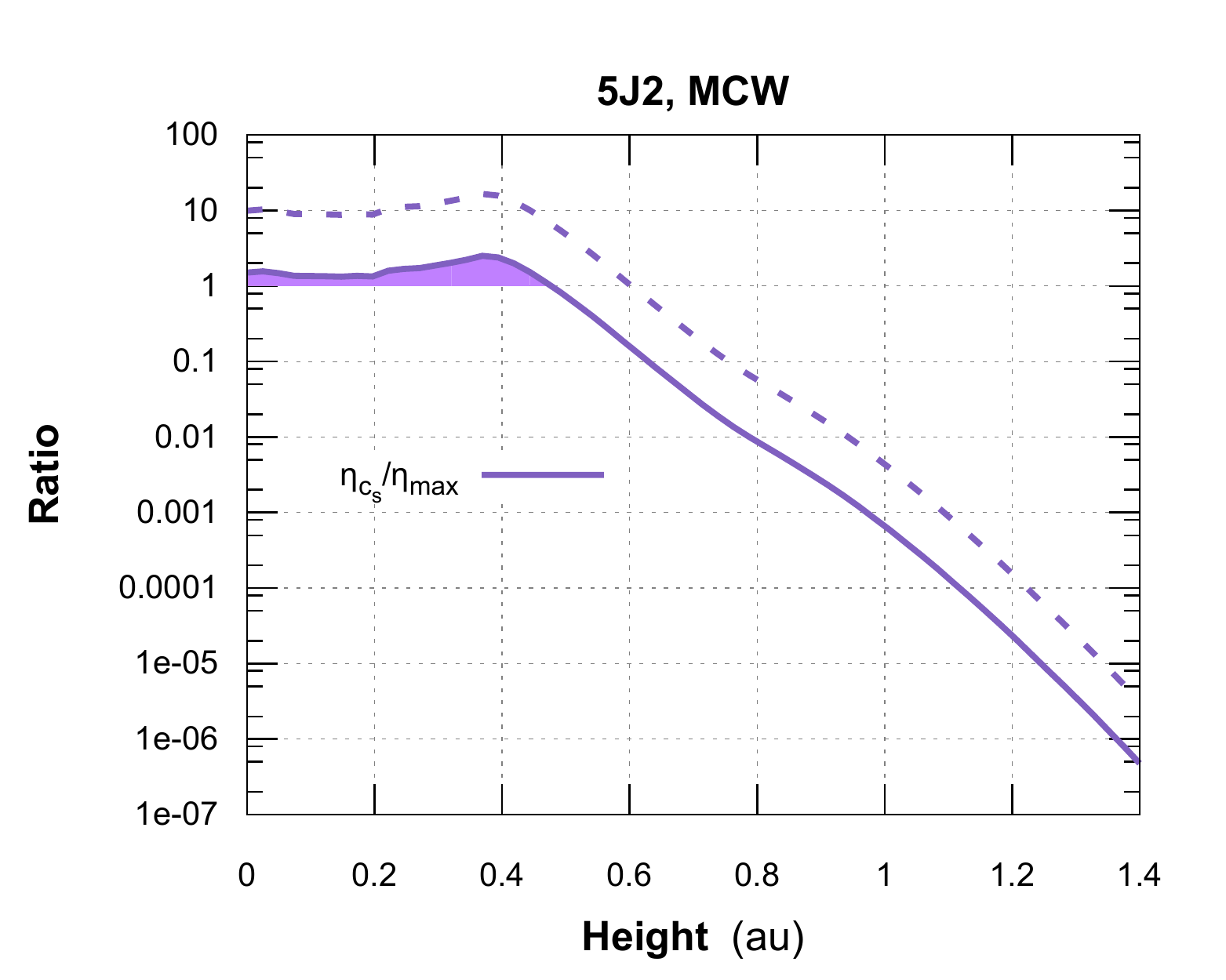}
\caption{Viability of the magneto-rotational turbulence (MRT),
  Hall-shear instability (HSI), and magneto-centrifugal wind (MCW)
  scenarios (top to bottom), in the models with a Saturn in a dusty
  disk (10S0, left column) and a Jupiter in a dust-depleted disk (5J2,
  right column).  Solid curves are for the mass flow rate found just
  outside the planet's orbit, dashed curves for the lower rate just
  inside the orbit.  Shading marks the portion of each solid curve
  that meets the constraints of eqs.~\ref{eq:etaPthreshold} (green),
  \ref{eq:betazthreshold} (red), \ref{eq:lHthreshold} (blue), and
  \ref{eq:windthreshold} (purple).  Darker sections of the blue curves
  show where the Hall diffusivity is positive.  The MRT scenario
  jointly meets its three constraints nowhere.  The HSI scenario is
  viable everywhere in the Saturn case, and in three distinct layers
  in the Jupiter case.  MCW yields diffusivities consistent with its
  own operation in the Saturn case in a layer of the disk atmosphere,
  and in the Jupiter case near the
  midplane.  \label{fig:diffusivities}}
\end{figure}

Profiles of the magnetic coupling through the column of material above
the planet are shown in figure~\ref{fig:diffusivities}.  The two cases
plotted here bracket the range of the full set of models.  The first,
10S0, has conditions least favorable for ionization: the planet lies
far from the X-ray source, the full dust abundance means both greater
X-ray optical depth and rapid recombination on grain surfaces, and the
relatively large gas column in the gap likewise makes recombination
quick.  The other case in figure~\ref{fig:diffusivities}, model 5J2,
is at the opposite extreme in all these respects, and has the highest
midplane ionization fraction of all our models.

Because the X-ray intensity rises with height above the planet, while
the recombination rate declines as the square of the density, the
ionization fraction rises with height.  In model 10S0, this means a
fairly well-coupled layer at intermediate heights, while still further
up, the declining density makes the ambipolar diffusivity dominant.
In model 5J2, lower gas densities and higher ionization rates mean the
fairly well-coupled layer reaches down to the midplane.  These trends
in the three non-ideal magnetic diffusivities with mass column are
similar to those in modeling of protostellar disks without gaps, for
example by \citet{2007Ap&SS.311...35W} and \citet{2011ApJ...739...50B,
  2014ApJ...791...72B}.

The coupling conditions shown in figure~\ref{fig:diffusivities} have
the following implications for the three magnetic scenarios.  From the
top left panel, we see that in the 10S0 model under the turbulent
scenario, the magnetic field near the midplane is weak enough to be
consistent with magneto-rotational turbulence (red shading), but the
diffusivity is too large for the instability to operate (the blue and
green curves are below unity).  Conversely in the atmosphere, the Hall
diffusivity is low enough for turbulence (blue shading), but the field
is too strong to allow it.  Ambipolar diffusion is an obstable to
turbulence everywhere (green curve).  The situation is
still-less-favorable for magneto-rotational turbulence in the 5J2
model (top right panel), as the magnetic field is too strong even at
the midplane.  The middle row of panels shows that the criterion for
Hall-shear instability is satisfied throughout the 10S0 model, and in
three distinct layers in the 5J2 model (blue shading).  The panels in
the bottom row show that a magneto-centrifugal wind can be launched
from a surface layer in the 10S0 case, and from the midplane up to
half a scale height or so in the 5J2 case.  The launching layers'
shallowness might suggest the wind does not reach escape speed before
ambipolar diffusion decouples it from the fields, ending magnetic
acceleration and collimation.  However, ambipolar diffusion can also
heat the wind \citep{1993ApJ...408..115S} enough to increase its
ionization beyond that provided by the X-rays.  Furthermore, much
greater ionization is expected high in the atmosphere where the
stellar far-ultraviolet photons are absorbed
\citep{2011ApJ...735....8P, 2011ApJ...739...78B}.  Since we treat
neither ambipolar heating nor FUV photons, our calculations are not
valid higher in the outflow.  Determining the fate of the wind would
furthermore require non-ideal MHD calculations spanning the gap and
the nearby disk, where launching conditions may be quite different
from the gap.


A note of caution is in order regarding timescales in the 5J2 model,
with the Jupiter-mass planet at 5~au in the dust-depleted disk.  Gas
flows from the gap edge to the planet in about 100~years, similar to
the time for the reaction network to reach equilibrium at the
midplane.  The ionization state in the gap will therefore be
intermediate between our local equilibrium values and conditions
upstream in the outer rim.  Since the denser rim material has slower
ionization and faster recombination, this will mean the gap is
more-poorly coupled to the magnetic fields than in
table~\ref{tab:viability} and figure~\ref{fig:diffusivities}.

\section{DISCUSSION AND CONCLUSIONS\label{sec:conclusions}}

We have carried out transfer calculations for X-rays emitted from the
corona of a young star into a surrounding protostellar disk containing
an embedded Saturn- or Jupiter-mass planet.  The distribution of the
material through which the X-rays pass is determined by the planet's
tides, which open a gap in the disk, and by the optical starlight,
whose heating effects determine the disk thickness under our
assumption of vertical hydrostatic equilibrium.  Some of the X-rays
scattered in the disk atmosphere reach and ionize the gas in the
planet's vicinity.  The ionization rates are comparable to or greater
than those produced in the interstellar medium by cosmic rays.
Ionization rates near the Jupiters are up to one order of magnitude
greater than at the same distance and column in the standard IG99 fit,
owing to the scattering from the gap's rims.  Ultraviolet radiation,
which we do not treat, could further ionize the planet's vicinity in
the Jupiter cases.  However the Saturn cases' greater column density,
which far exceeds ultraviolet photons' absorption depth, means there
are few options for raising the ionization rate near the planet above
that provided by the X-rays.

The ionization rates depend mostly on the shape of the gap and the
column of material within.  The magnetic coupling results summarized
next depend also on the recombination chemistry and magnetic field
strength and orientation, for which the ranges of possibilities are
wide.  The coupling results should therefore be considered less
certain.

We compute the equilibrium ionization state of the material near the
planet using a simplified chemical reaction network including
representative molecular and metal ions, and recombination on grain
surfaces.  From the populations of charged species, we compute the
Ohmic, Hall, and ambipolar diffusivities under three scenarios for the
magnetic fields: the accretion flow is driven by either (1)
magneto-rotational turbulence, (2) Hall-shear instability, or (3) a
magneto-centrifugal wind.

In all cases, the diffusivity is too high for magneto-rotational
turbulence to transport angular momentum in the gas near the planet.
Ambipolar diffusion decouples the magnetic fields from the disk's
neutral component over the length and time scales that would be
associated with the turbulence.  Hall-shear instability, in contrast,
can drive the accretion flow in almost all cases: the field strengths
needed for HSI to produce the assumed mass flow rates yield Hall
diffusivities great enough for the HSI to operate.  This fails only in
the models with a Jupiter in a dust-depleted disk.  A scenario with a
magneto-centrifugal wind also appears viable in the less-diffusive
cases we considered, especially if ultraviolet ionization or ambipolar
heating sustains coupling at heights above where the X-rays are
important.  In particular, a wind can begin near the planet in all the
dust-depleted scenarios.

Overall, the X-rays leave material near the planet marginally ionized,
so that magnetic forces can alter the material's distribution and
impact the planet's growth and orbital migration, yet
magneto-rotational turbulence is unlikely.

While each of our models is static, together they suggest the
following evolutionary sequence.  A young planetary system develops as
the disk's solid material is incorporated into rocks, planetesimals,
and planets.  The planets grow by accreting much of the disk's solids
and a little of its gas, so the ambient dust-to-gas ratio falls with
time.  Planets growing from Saturn- to Jupiter-mass pass from
conditions resembling our 10S0 model to those more similar to the 5J2
model.  The magnetic coupling regime in the deepening gap becomes less
favorable to Hall shear, and more favorable to a magneto-centrifugal
wind, so the field realigns.  Much of the toroidal component is
expelled, while the vertical flux is retained due to the constraints
of the surrounding disk and wind.  As a result, the mass flow onto the
planet slows.

We have considered disks with a single planet.  A second planet,
interior to the first, could open a gap of its own, with the rim
casting an X-ray shadow if the disk is not too flared.  An inclined
interior planet also could warp the disk nearer the star, creating a
lighthouse effect where more X-rays reach the planet alternately from
one side of the disk and then the other.

Although we have neglected the gradient term in the mass flow--stress
relation (eq.~\ref{eq:mdot}) when estimating the field strengths
needed to drive the flows, this term could dominate around
planet-opened gaps, where the diffusivities and hence the magnetic
fields may change over distances comparable to the density scale
height.  In the simple model disk we used, the mass flow rate changes
across the planet's orbit.  If conditions vary sharply near the
planet, the diffusivity will likewise have steep gradients.  Magnetic
fields have been shown to evolve toward a more nearly uniform radial
profile than the gas in ideal-MHD, unstratified MRT calculations
\citep{2013ApJ...768..143Z}.  Evaluating the magnetic gradients in a
more diffusive environment with radial and vertical structure in the
ionization state will require further detailed MHD calculations.
While 3-D MHD models of disks with planet-opened gaps have been
constructed including the Ohmic diffusivity
\citep{2013ApJ...779...59G}, it now seems the ambipolar and Hall terms
are more important still.

\acknowledgements We are grateful to Barbara Ercolano for advice on
the X-ray opacities, Wilhelm Kley and Stephen Lubow for advice on
planet-opened gaps' depths, and Satoshi Okuzumi for providing his
subroutine implementing the Bethell \& Bergin cross-sections.  Lynne
Hillenbrand's sponsorship of SYK at Caltech enabled this work to
begin.  The research was carried out in part at the Jet Propulsion
Laboratory, California Institute of Technology, under a contract with
the National Aeronautics and Space Administration, and with the
support of the NASA Origins of Solar Systems program through grant
13-OSS13-0114 and Exoplanets Research Program through grant
17-XRP17\_2-0081.  Copyright 2019 California Institute of Technology.
Government sponsorship acknowledged.

\bibliographystyle{apj}
\bibliography{ysodisk}

\begin{thebibliography}{43}
\expandafter\ifx\csname natexlab\endcsname\relax\def\natexlab#1{#1}\fi

\bibitem[{{Bai}(2011)}]{2011ApJ...739...50B}
{Bai}, X.-N. 2011, \apj, 739, 50

\bibitem[{{Bai}(2014)}]{2014ApJ...791...72B}
---. 2014, \apj, 791, 72

\bibitem[{{Bai} \& {Goodman}(2009)}]{2009ApJ...701..737B}
{Bai}, X.-N., \& {Goodman}, J. 2009, \apj, 701, 737

\bibitem[{{Baruteau} {et~al.}(2014){Baruteau}, {Crida}, {Paardekooper},
  {Masset}, {Guilet}, {Bitsch}, {Nelson}, {Kley}, \&
  {Papaloizou}}]{2014prpl.conf..667B}
{Baruteau}, C., {et~al.} 2014, Protostars and Planets VI, 667

\bibitem[{{Bethell} \& {Bergin}(2011{\natexlab{a}})}]{2011ApJ...740....7B}
{Bethell}, T.~J., \& {Bergin}, E.~A. 2011{\natexlab{a}}, \apj, 740, 7

\bibitem[{{Bethell} \& {Bergin}(2011{\natexlab{b}})}]{2011ApJ...739...78B}
---. 2011{\natexlab{b}}, \apj, 739, 78

\bibitem[{{Bjorkman} \& {Wood}(2001)}]{2001ApJ...554..615B}
{Bjorkman}, J.~E., \& {Wood}, K. 2001, \apj, 554, 615

\bibitem[{{Cleeves} {et~al.}(2013){Cleeves}, {Adams}, \&
  {Bergin}}]{2013apj...772....5i}
{Cleeves}, L.~I., {Adams}, F.~C., \& {Bergin}, E.~A. 2013, \apj, 772, 5

\bibitem[{{Cleeves} {et~al.}(2015){Cleeves}, {Bergin}, {Qi}, {Adams}, \&
  {{\"O}berg}}]{2015ApJ...799..204C}
{Cleeves}, L.~I., {Bergin}, E.~A., {Qi}, C., {Adams}, F.~C., \& {{\"O}berg},
  K.~I. 2015, \apj, 799, 204

\bibitem[{{D'Antona} \& {Mazzitelli}(1994)}]{1994ApJS...90..467D}
{D'Antona}, F., \& {Mazzitelli}, I. 1994, \apjs, 90, 467

\bibitem[{{D{\"u}rmann} \& {Kley}(2015)}]{2015A&A...574A..52D}
{D{\"u}rmann}, C., \& {Kley}, W. 2015, \aap, 574, A52

\bibitem[{{Ercolano} \& {Glassgold}(2013)}]{2013MNRAS.436.3446E}
{Ercolano}, B., \& {Glassgold}, A.~E. 2013, \mnras, 436, 3446

\bibitem[{{Feigelson} \& {Montmerle}(1999)}]{1999ARA&A..37..363F}
{Feigelson}, E.~D., \& {Montmerle}, T. 1999, \araa, 37, 363

\bibitem[{{Fung} \& {Chiang}(2016)}]{2016ApJ...832..105F}
{Fung}, J., \& {Chiang}, E. 2016, \apj, 832, 105

\bibitem[{{Fung} {et~al.}(2014){Fung}, {Shi}, \&
  {Chiang}}]{2014ApJ...782...88F}
{Fung}, J., {Shi}, J.-M., \& {Chiang}, E. 2014, \apj, 782, 88

\bibitem[{{Garmire} {et~al.}(2000){Garmire}, {Feigelson}, {Broos},
  {Hillenbrand}, {Pravdo}, {Townsley}, \& {Tsuboi}}]{2000AJ....120.1426G}
{Garmire}, G., {Feigelson}, E.~D., {Broos}, P., {Hillenbrand}, L.~A., {Pravdo},
  S.~H., {Townsley}, L., \& {Tsuboi}, Y. 2000, \aj, 120, 1426

\bibitem[{{Glassgold} {et~al.}(1997){Glassgold}, {Najita}, \&
  {Igea}}]{1997ApJ...480..344G}
{Glassgold}, A.~E., {Najita}, J., \& {Igea}, J. 1997, \apj, 480, 344

\bibitem[{{Gressel} {et~al.}(2013){Gressel}, {Nelson}, {Turner}, \&
  {Ziegler}}]{2013ApJ...779...59G}
{Gressel}, O., {Nelson}, R.~P., {Turner}, N.~J., \& {Ziegler}, U. 2013, \apj,
  779, 59

\bibitem[{{Hawley} {et~al.}(1995){Hawley}, {Gammie}, \&
  {Balbus}}]{1995ApJ...440..742H}
{Hawley}, J.~F., {Gammie}, C.~F., \& {Balbus}, S.~A. 1995, \apj, 440, 742

\bibitem[{{Hayashi}(1981)}]{1981PThPS..70...35H}
{Hayashi}, C. 1981, Progress of Theoretical Physics Supplement, 70, 35

\bibitem[{{Igea} \& {Glassgold}(1999)}]{1999ApJ...518..848I}
{Igea}, J., \& {Glassgold}, A.~E. 1999, \apj, 518, 848

\bibitem[{{Ilgner} \& {Nelson}(2006)}]{2006A&A...445..205I}
{Ilgner}, M., \& {Nelson}, R.~P. 2006, \aap, 445, 205

\bibitem[{{Karzas} \& {Latter}(1961)}]{1961ApJS....6..167K}
{Karzas}, W.~J., \& {Latter}, R. 1961, \apjs, 6, 167

\bibitem[{{Keith} \& {Wardle}(2015)}]{2015MNRAS.451.1104K}
{Keith}, S.~L., \& {Wardle}, M. 2015, \mnras, 451, 1104

\bibitem[{{Lesur} {et~al.}(2014){Lesur}, {Kunz}, \&
  {Fromang}}]{2014A&A...566A..56L}
{Lesur}, G., {Kunz}, M.~W., \& {Fromang}, S. 2014, \aap, 566, A56

\bibitem[{{Lin} \& {Papaloizou}(1986)}]{1986ApJ...309..846L}
{Lin}, D.~N.~C., \& {Papaloizou}, J. 1986, \apj, 309, 846

\bibitem[{{Lubow} \& {D'Angelo}(2006)}]{2006ApJ...641..526L}
{Lubow}, S.~H., \& {D'Angelo}, G. 2006, \apj, 641, 526

\bibitem[{{Lucy}(1999)}]{1999A&A...344..282L}
{Lucy}, L.~B. 1999, \aap, 344, 282

\bibitem[{{Okuzumi} \& {Hirose}(2011)}]{2011ApJ...742...65O}
{Okuzumi}, S., \& {Hirose}, S. 2011, \apj, 742, 65

\bibitem[{{Perez-Becker} \& {Chiang}(2011)}]{2011ApJ...735....8P}
{Perez-Becker}, D., \& {Chiang}, E. 2011, \apj, 735, 8

\bibitem[{{Preibisch} {et~al.}(1993){Preibisch}, {Ossenkopf}, {Yorke}, \&
  {Henning}}]{1993A&A...279..577P}
{Preibisch}, T., {Ossenkopf}, V., {Yorke}, H.~W., \& {Henning}, T. 1993, \aap,
  279, 577

\bibitem[{{Safier}(1993)}]{1993ApJ...408..115S}
{Safier}, P.~N. 1993, \apj, 408, 115

\bibitem[{{Sano} \& {Inutsuka}(2001)}]{2001ApJ...561L.179S}
{Sano}, T., \& {Inutsuka}, S.-i. 2001, \apjl, 561, L179

\bibitem[{{Sano} {et~al.}(2004){Sano}, {Inutsuka}, {Turner}, \&
  {Stone}}]{2004ApJ...605..321S}
{Sano}, T., {Inutsuka}, S.-i., {Turner}, N.~J., \& {Stone}, J.~M. 2004, \apj,
  605, 321

\bibitem[{{Sano} \& {Stone}(2002)}]{2002ApJ...570..314S}
{Sano}, T., \& {Stone}, J.~M. 2002, \apj, 570, 314

\bibitem[{{Siess} {et~al.}(2000){Siess}, {Dufour}, \&
  {Forestini}}]{2000A&A...358..593S}
{Siess}, L., {Dufour}, E., \& {Forestini}, M. 2000, \aap, 358, 593

\bibitem[{{Turner} {et~al.}(2012){Turner}, {Choukroun}, {Castillo-Rogez}, \&
  {Bryden}}]{2012ApJ...748...92T}
{Turner}, N.~J., {Choukroun}, M., {Castillo-Rogez}, J., \& {Bryden}, G. 2012,
  \apj, 748, 92

\bibitem[{{Turner} \& {Sano}(2008)}]{2008ApJ...679L.131T}
{Turner}, N.~J., \& {Sano}, T. 2008, \apjl, 679, L131

\bibitem[{{Turner} {et~al.}(2007){Turner}, {Sano}, \&
  {Dziourkevitch}}]{2007ApJ...659..729T}
{Turner}, N.~J., {Sano}, T., \& {Dziourkevitch}, N. 2007, \apj, 659, 729

\bibitem[{{Umebayashi} \& {Nakano}(2009)}]{2009ApJ...690...69U}
{Umebayashi}, T., \& {Nakano}, T. 2009, \apj, 690, 69

\bibitem[{{Wardle}(2007)}]{2007Ap&SS.311...35W}
{Wardle}, M. 2007, \apss, 311, 35

\bibitem[{{Weidenschilling}(1977)}]{1977Ap&SS..51..153W}
{Weidenschilling}, S.~J. 1977, \apss, 51, 153

\bibitem[{{Zhu} {et~al.}(2013){Zhu}, {Stone}, \&
  {Rafikov}}]{2013ApJ...768..143Z}
{Zhu}, Z., {Stone}, J.~M., \& {Rafikov}, R.~R. 2013, \apj, 768, 143

\end{thebibliography}

\end{document}